\documentclass[a4paper,usenatbib,useAMS]{mnras}
\usepackage{subfigure}
\usepackage{mathtools}
\usepackage{graphicx}	% Including figure files
\usepackage{amsmath}	% Advanced maths commands
\usepackage{amssymb}	% Extra maths symbols
\usepackage{cancel}
\usepackage[usenames]{color}
\usepackage{hyperref}
\usepackage{cleveref}
\usepackage{mathrsfs}
   \usepackage[british]{babel}             % British English hyphenation
   \usepackage{newtxtext}                  % Good fonts
   \usepackage[T1]{fontenc}                % Good font encoding
   \usepackage{newtxtext}
\crefname{figure}{Fig.}{Figs.}
\newcommand{\equ}[1]{\cref{eq:#1}} %%     eq.~(\ref{eq:#1})}
\newcommand{\equs}[2]{\cref{eq:#1,eq:#2}} %%     eq.~(\ref{eq:#1})}
\newcommand{\se}[1]{\S\ref{sec:#1}}
\newcommand{\figref}[1]{\cref{fig:#1}}%%   fig. \ref{fig:#1}}

\usepackage{pifont}% http://ctan.org/pkg/pifont

\newcommand{\ddt}[1]{\frac{\partial{#1}}{\partial t}}
\newcommand{\cs}{c_\mathrm{s}}
\newcommand{\tturb}{t_\mathrm{turb}}
\newcommand{\tff}{t_\mathrm{ff}}
\newcommand{\tsound}{t_\mathrm{sound}}
\newcommand{\mach}{\mathcal{M}}

\newcommand{\bturb}{B_\mathrm{turb}}
\newcommand{\va}{v_\mathrm{A}}
\usepackage{bm}
\newcommand{\taub}{\bm{\tau}}

\usepackage[normalem]{ulem}

\definecolor{darkgreen}{rgb}{0.13, 0.55, 0.13}

\def\geff{\gamma_{\rm eff}}
\def\ekin{\epsilon_{\rm kin}}
\def\sigsat{\sigma_{\rm sat}}

\newcommand{\figtop}[1]{{\cref{fig:#1} ({\it top panel})}}

\def\top{{\it top panel}}
\def\bot{{\it bottom panel}}

\begin{document}
\label{firstpage}
\pagerange{\pageref{firstpage}--\pageref{lastpage}}
\title[Turbulent Compression of Magnetized Gas]{Compression of turbulent magnetized gas in Giant Molecular Clouds}
\author[Y. Birnboim, C. Federrath \& M. Krumholz]{Yuval Birnboim$^\text{1,2}$\thanks{Contact email: yuval@phys.huji.ac.il}, Christoph Federrath$^\text{1}$ \& Mark Krumholz$^\text{1}$\\$^\text{1}$Research School of Astronomy \& Astrophysics, Australian National University, Canberra, ACT, Australia\\$^\text{2}$Racah Institute of Physics, The Hebrew University, Jerusalem 91904,  Israel\\}

%\author{Yuval Birnboim}
%\affiliation{Research School of Astronomy \& Astrophysics, Australian National University, Canberra%, ACT, Australia}
%\affiliation{Racah Institute of Physics, The Hebrew University, Jerusalem 91904,  Israel}

%\author{Christoph Federrath}
%\affiliation{Research School of Astronomy \& Astrophysics, Australian National University, Canberra, ACT, Australia}

%\author{Mark Krumholz}
%\affiliation{Research School of Astronomy \& Astrophysics, Australian National University, Canberra, ACT, Australia}

%%\correspondingauthor{Yuval Birnboim}
%%\email{yuval@phys.huji.ac.il}

\date{Last updated 2017 May 23; in original form 2017 May 23}
\pubyear{2017}

\maketitle

\begin{abstract}
Interstellar gas clouds are often both highly magnetized and supersonically turbulent, with velocity dispersions set by a competition between driving and dissipation. This balance has been studied extensively in the context of gases with constant mean density. However, many astrophysical systems are contracting under the influence of external pressure or gravity, and the balance between driving and dissipation in a contracting, magnetized medium has yet to be studied. In this paper we present three-dimensional (3D) magnetohydrodynamic (MHD) simulations of compression in a turbulent, magnetized medium that resembles the physical conditions inside molecular clouds. We find that in some circumstances the combination of compression and magnetic fields leads to a rate of turbulent dissipation far less than that observed in non-magnetized gas, or in non-compressing magnetized gas. As a result, a compressing, magnetized gas reaches an equilibrium velocity dispersion much greater than would be expected for either the hydrodynamic or the non-compressing case. We use the simulation results to construct an analytic model that gives an effective equation of state for a coarse-grained parcel of the gas, in the form of an ideal equation of state with a polytropic index that depends on the dissipation and energy transfer rates between the magnetic and turbulent components. We argue that the reduced dissipation rate and larger equilibrium velocity dispersion has important implications for the driving and maintenance of turbulence in molecular clouds, and for the rates of chemical and radiative processes that are sensitive to shocks and dissipation.
\end{abstract}


% SECTION: introduction
\section{Introduction}
Magnetized plasma is ubiquitous in astrophysical systems. Particularly, gas in the interstellar medium (ISM) is observed to be magnetized, and a large fraction of its energy content is in the form of magnetic fields \citep[see][and references within]{beck96,ferriere01,Federrath2016jpp}. The magnetic field in the ISM of disk galaxies consists of an ordered rotating component on galactic disk scales that is consistent with slow winding of the magnetic field via macroscopic dynamo processes, and small-scale magnetic fields that are generated by the winding of the magnetic field via turbulent dynamo processes \citep{BrandenburgSubramanian2005,BrandenburgSokoloffSubramanian2012}. For the Galaxy, the values of the two components are comparable, with typical values of $2$--$5\,\mu\mathrm{G}$.

When portions of this magnetized fluid are subject to rapid radiative cooling, for example in molecular clouds, the result is a highly supersonic, strongly magnetized flow. Within such a flow, the velocity dispersion is dictated by the balance between driving and dissipation processes. This balance, particularly the dissipation part of it, has been studied extensively for both non-magnetized and magnetized flows in the context of periodic boxes with constant mean density \citep[e.g.][]{maclow98, StoneOstrikerGammie1998, OstrikerGammieStone1999, PadoanNordlund1999, maclow99, KritsukEtAl2007, Lemaster09a}. The general result from these simulations is that the turbulence decays on a timescale comparable to a large eddy turnaround time, and that the rates of decay are not substantially altered by the presence or absence of a magnetic field.

The problem of the balance between driving and decay for magnetized turbulence is most acute in molecular clouds. Since these have linewidths indicating the presence of supersonic flow, the fast dissipation of turbulence found by these simulations necessitates a mechanism to reinject the energy equally quickly. A number of candidates have been proposed, including internal feedback from H~\textsc{ii} regions \citep{Matzner02a, Krumholz06d, Goldbaum11a} or protostellar outflows \citep{Li06b, Nakamura07a, Wang10a, FederrathEtAl2014}, driving of turbulence by ongoing accretion \citep{Klessen10a, Goldbaum11a, Lee16b} or gravitational contraction on small scales \citep{FederrathSurSchleicherBanerjeeKlessen2011,SurEtAl2012}, thermal instability driving \mbox{\citep{Koyama02a, Hennebelle06a}} and injection of energy from external supernova shocks \citep{MacLowKlessen2004, Padoan16b, Padoan16a, PanEtAl2016}. Alternately, it is possible that the linewidths do not reflect turbulent motion at all, and instead indicate global gravitational collapse \citep{Ballesteros-Paredes11a, Zamora-Aviles14a}. Each of these proposals, however, faces challenges -- internal feedback must maintain large linewidths without destroying the clouds in which they occur, driving by accretion faces the problem of what happens when the accretion eventually ends, thermal instability seems unlikely to be a viable mechanism in molecule-dominated galaxies that lack a significant warm phase, and external driving requires efficient coupling between the low density external medium and the dense clouds. The view that clouds are in global collapse is hard to reconcile with the observed very low rates of star formation found even in gas at densities $\gtrsim 10^5$ cm$^{-3}$ \citep{Krumholz07e, Krumholz12a, FederrathKlessen2012, PadoanEtAl2014, Evans14a, Usero15a, Salim15a, Vutisalchavakul16a, Heyer16a}.

The problem of the persistence of turbulence in molecular clouds is significantly eased if gravitational compression is able to pump energy into turbulent motion, since this would provide a mechanism to both power the turbulence and slow the collapse. The phenomenon has been explored for non-magnetized flows by \citet{robertson12}. In their work, an initially-turbulent gas is compressed in a scale-free manner by renormalizing the thermodynamic variables according to the expected values from a uniform collapse. As gas is compressed, the amplitude of the velocity field increases because the compression does $P\, dV$ work against the kinetic pressure. On the other hand, the typical size of the eddies is reduced by the compression, and this accelerates the decay of turbulence. Depending on the compression rate, one process or the other dominates, and the turbulence either increases or decays. Qualitatively, the results are consistent with what one would have derived by naively equating the rate of $P\, dV$ work with a decay timescale of $\sim 1$ eddy turnover time derived from non-compressing driven turbulence simulations: the turbulence is amplified when the box compression time is short compared to the eddy turnover time, and decays if the converse holds. However, \citet{robertson12} did not include magnetic fields in their simulations, and we know that all clouds in the ISM are magnetized to a level that corresponds to a near equipartition between turbulent and magnetic energy densities \citep[][and references therein]{Crutcher12a}.

In this work we seek to determine whether the result by \citet{robertson12} is altered in the presence of a magnetic field. We have already noted that, in driven turbulence simulations, magnetic fields make no qualitative difference. However, driving turbulence by global compression is qualitatively different than direct driving of the gas. In the first case, the scaling relations of velocity and distance for global compression enhance all modes similarly, and some modes, for which the dissipation is faster and that are not replenished quickly enough by a turbulent cascade, can decay and disappear. In the second, the turbulence forcing arbitrarily sets the geometry of the flow and phases of the various modes, preventing the flow from achieving a more relaxed state.

A magnetic field might change the situation for a compressing flow in two ways. First, a magnetic field and gas motions can exchange energy via a turbulent dynamo \citep{Kazantsev1968,Subramanian1997,Subramanian1999,BrandenburgSubramanian2005,SchleicherEtAl2010,SchoberEtAl2012PRE2,SchoberEtAl2012,SchoberEtAl2012PRE,BovinoEtAl2013,schober15}. For driven turbulence, the amount of energy stored in the dynamo is limited by the back reaction of the Lorentz forces on the gas, \citep{federrath11, Federrath2016jpp}, and as a result the energy stored in the magnetic field is always subdominant compared to the turbulence. However, gravitational compression will amplify magnetic fields differently than gas motions \citep{SurEtAl2010,FederrathSurSchleicherBanerjeeKlessen2011,SurEtAl2012}, potentially leading to magnetic-turbulent interactions not found in driven, non-compressing boxes. Second, magnetic fields will impose anisotropy on the flow, and anisotropic turbulence shows a different cascade pattern and a different decay rate than isotropic turbulence \citep{Cho02a, Cho03a, Hansen11a}.

In this paper we examine the effect of magnetic fields on global compression of turbulent gas in idealized 3D MHD simulations. We distinguish between cases of zero net magnetic flux and cases with non-zero net flux of various amplitudes. This is of theoretical importance because a magnetic field with finite flux increases monotonically as gas contracts, and of practical interest because fields with non-zero net flux are likely present in proto-GMCs \citep{LiHenning2011,LiEtAl2011,PillaiEtAl2015}. Rather than introducing cooling, we assume that the gas is isothermal, which is a reasonable approximation for GMCs over a wide range of densities. We describe our setup in \se{sims} and our simulated results in \se{results}. We then construct an analytic prediction for the effective equation of state of a system with mixed thermal, kinetic and magnetic pressure components (\se{model}), use some of the physical insights to further analyze the dissipation in the simulations and compare our predictions to the simulations (\se{comparison}). In \se{discussion} we discuss possible implication of our results to ISM and GMCs, and in \se{conclusions} we summarize and conclude.

% SECTION Simulations (methods)
\section{Simulations} \label{sec:sims}

\subsection{The FLASH code}

We use a modified version of the grid-based code FLASH \citep{FryxellEtAl2000,DubeyEtAl2008} (\url{http://www.flash.uchicago.edu/site/flashcode/}) to solve the three-dimensional (3D), compressible, ideal magnetohydrodynamical (MHD) equations,
\begin{align}
& \ddt\,\rho + \nabla\cdot\left(\rho \mathbf{v}\right)=0, \label{eq:mhd1} \\
& \ddt\!\left(\rho \mathbf{v}\right) + \nabla\cdot\left(\rho \mathbf{v}\!\otimes\!\mathbf{v} - \frac{1}{4\pi}\mathbf{B}\!\otimes\!\mathbf{B}\right) + \nabla P_\mathrm{tot} = 0, \label{eq:mhd2} \\
& \ddt\,e + \nabla\cdot\left[\left(e+P_\mathrm{tot}\right)\mathbf{v} - \frac{1}{4\pi}\left(\mathbf{B}\cdot\mathbf{v}\right)\mathbf{B}\right] = 0, \label{eq:mhd3} \\
& \ddt\,\mathbf{B} - \nabla\times\left(\mathbf{v}\times\mathbf{B}\right) = 0,\quad\nabla\cdot\mathbf{B} = 0. \label{eq:mhd4}
\end{align}
Here, $\rho$, $\mathbf{v}$, $P_\mathrm{tot}=P_\mathrm{th}+ (1/8\pi)\left|\mathbf{B}\right|^2$, $\mathbf{B}$, and $e=\rho \epsilon_\mathrm{int} + (1/2)\rho\left|\mathbf{v}\right|^2 + (1/8\pi)\left|\mathbf{B}\right|^2$ denote the gas density, velocity, pressure (thermal plus magnetic), magnetic field, and total energy density (internal, plus kinetic, plus magnetic), respectively. The MHD equations are closed with a quasi-isothermal equation of state (EoS), $P_\mathrm{th}=(\gamma-1)\rho\epsilon_\mathrm{int}$, where we set $\gamma=1.00001$. Using this setting we model a gas with an extremely high number of degrees of freedom, $f=2/(\gamma-1)\sim2\times10^5$, effectively resulting in a gas that is isothermal. This is a standard procedure to obtain a quasi-isothermal EoS and results in the same thermodynamic response of the gas as a polytropic EoS, $P_\mathrm{th}\propto\rho^\Gamma$ with $\Gamma=1$ \citep{FederrathEtAl2014}. The practical reason for choosing the simple ideal gas EoS with $\gamma=1.00001$ is to keep track of how much energy is dissipated by the turbulence, i.e., we solve the energy equation (\ref{eq:mhd3}) and record the change in energy every timestep.

The system of ideal MHD equations~(\ref{eq:mhd1}--\ref{eq:mhd4}) are solved with the robust HLL3R Riemann scheme by \citet{WaaganFederrathKlingenberg2011}, based on previous developments in applied mathematics to preserve positive density and pressure by construction \citep{BouchutKlingenbergWaagan2007,KlingenbergSchmidtWaagan2007,Waagan2009,BouchutKlingenbergWaagan2010}. For our particular simulations, the magnetic field is shown to remain divergence free to a reasonable degree (Appendix~\se{divB})

\subsection{Numerical scheme for solving the MHD equations in an expanding or contracting coordinate system}

The cosmology unit in FLASH allows one to use any hydrodynamics solver written for a non-expanding universe to work unmodified in a cosmological context. This is achieved by solving the MHD equations in the co-moving reference frame and accounting for the additional terms in the MHD equations that appear due to the expansion or contraction of the system. All calculations are assumed to take place in co-moving coordinates $\mathbf{x}$ = $\mathbf{r}/a$, where $\mathbf{r}$ is the physical (proper) position vector and $a(t)$ is the time-dependent cosmological scale factor. When transforming the MHD equations to the co-moving frame, the spatial derivative transforms as $\nabla_\mathbf{x} = a \nabla_\mathbf{r}$ and the time derivative transforms as $(\partial/\partial t)_\mathbf{x} = (\partial/\partial t)_\mathbf{r} + H\mathbf{r}\cdot\nabla_\mathbf{r}$, where the Hubble constant is defined as $H=\dot{a}/a$. The physical (proper) velocity is given as $\tilde{\mathbf{v}} = H\mathbf{r} + a\dot{\mathbf{x}}$, where the first term is the Hubble flow and the second term contains the co-moving velocity $\mathbf{v}=\dot{\mathbf{x}}$.

Using these relations between the physical and co-moving derivatives in addition to the following transformations from physical (with tilde) to co-moving hydrodynamical quantities (without tilde),
\begin{align}
\rho &= a^3 \tilde{\rho}, \\
\mathbf{B} &= a^{1/2} \tilde{\mathbf{B}}, \\
P_\mathrm{tot} &= a \tilde{P}_\mathrm{tot}, \\
e &= a \tilde{e}, \\
\epsilon_\mathrm{int} &= a^{-2} \tilde{\epsilon}_\mathrm{int},
\end{align}
the MHD equations in co-moving coordinates have exactly the same form as Equations~(\ref{eq:mhd1}--\ref{eq:mhd4}) with additional Hubble source terms on the right-hand sides of the momentum, energy and induction equations:
\begin{align}
& \ddt\,\rho + \nabla\cdot\left(\rho \mathbf{v}\right)=0, \label{eq:mhd1h} \\
& \! \begin{multlined}
	\ddt\!\left(\rho \mathbf{v}\right) + \nabla\cdot\left(\rho \mathbf{v}\!\otimes\!\mathbf{v} - \frac{1}{4\pi}\mathbf{B}\!\otimes\!\mathbf{B}\right) + \nabla P_\mathrm{tot} = {} \\
	-2H\rho\mathbf{v},
\end{multlined} \label{eq:mhd2h} \\
& \! \begin{multlined}
	\ddt\,e + \nabla\cdot\left[\left(e+P_\mathrm{tot}\right)\mathbf{v} - \frac{1}{4\pi}\left(\mathbf{B}\cdot\mathbf{v}\right)\mathbf{B}\right] = {} \\
	\phantom{..........................}-H\left[(3\gamma-1)\rho\epsilon_\mathrm{int}+2\rho\mathbf{v}\cdot\mathbf{v}\right],
\end{multlined} \label{eq:mhd3h} \\
& \ddt\,\mathbf{B} - \nabla\times\left(\mathbf{v}\times\mathbf{B}\right) = -\frac{3}{2}H\mathbf{B},\quad\nabla\cdot\mathbf{B} = 0. \label{eq:mhd4h}
\end{align}
Note that we have changed all time and space derivatives in these equations to the co-moving frame, i.e., $\partial/\partial t \equiv (\partial/\partial t)_\mathbf{x}$ and $\nabla\equiv\nabla_\mathbf{x}$.

Since the form of these equations is identical to the conservation Equations~(\ref{eq:mhd1}--\ref{eq:mhd4}) without the Hubble source terms, we can use any existing hydrodynamical scheme to solve this set of equations in the co-moving frame. In order to account for the Hubble source terms on the right-hand side of these equations, we use an operator-splitting approach, where the co-moving hydrodynamical variables are modified in each time step (after the hydro step) to account for the source terms.

First we note that the mass continuity equation is unchanged between physical and co-moving coordinates. The momentum equation has the Hubble source term $-2H\rho\mathbf{v}$. Expanding the co-moving momentum equation (\ref{eq:mhd2h}) with respect to the change in $a$, we find $\cancel{\dot{\rho}v}+\rho\dot{v}+\cancel{\nabla(\dots)}=-2H\rho v$, where $\dot{\rho}=(d\rho/da)(da/dt)=0$ because $d\rho/da=0$, and any spatial derivatives cancel, because $a$ does not depend on space. This leaves us with the simple differential equation, $\dot{v}/v=-2\dot{a}/a$, for which the solution is $v'=v(a/a')^2$, where $v$ and $a$ are the velocity and scale factor before accounting for the Hubble term (i.e., before the hydro step) and $v'$ and $a'=a(t+\Delta t)$ are the velocity and scale factor after the current time step $\Delta t$. An analogous correction has to be made in the co-moving energy equation to account for the Hubble source term, i.e., $\epsilon_\mathrm{int}'=\epsilon_\mathrm{int}(a/a')^{3\gamma-1}$.

These procedures to account for the Hubble flow in pure hydrodynamics (without magnetic fields) were already implemented in the cosmology module of the public version of FLASH. However, MHD was not supported. Here we implemented the necessary modifications of the induction equation with the Hubble source term $-(3/2)H\mathbf{B}$ in Equation~(\ref{eq:mhd4h}), which requires a modification of the co-moving magnetic field with $B'=B(a/a')^{3/2}$, analogous to the operator-split corrections for the velocity and energy explained in the previous paragraph.

\subsection{Initial driving of turbulence} \label{sec:driving}

\begin{table*}
\caption{List of simulation parameters at the beginning of the contraction phase.}
\label{tab:sims}
\def\arraystretch{1.5}
\begin{tabular*}{\linewidth}{@{\extracolsep{\fill} }lcccccc}
\hline
\hline
Simulation model & $\langle B_z\rangle$ & $\bturb$ & $H$ &$\mach$ & $\sigma_{\rm sat}(\%)$ &$N_\mathrm{res}^3$ \\
\hline
HD & $0$ & $0$ & $-200$ & $9.4$ & $0$ & $512^3$ \\
HD-H1 & $0$ & $0$ & $-1$ & $9.4$ & $0$ & $512^3$ \\
\hline
noGF-Medium & $0$ & $9.2$ & $-200$ &$9.8$ & $8$ &$512^3$ \\
noGF-Strong & $0$ & $24$ & $-200$ &$9.1$ & $56$ & $512^3$ \\
noGF-Strong-LR & $0$ & $21$ & $-200$ &$9.5$ & $40$ & $256^3$ \\
noGF-Strong-H1 & $0$ & $24$ & $-1$ &$9.1$ & $56$ & $512^3$ \\
\hline
GF-Weak & $0.35$ & $5.5$ & $-200$ &$9.3$ & $3$ & $512^3$ \\
GF-Medium & $3.5$ & $18$ & $-200$ &$9.5$ & $30$ & $512^3$ \\
GF-Medium-H1 & $3.5$ & $18$ & $-1$ &$9.5$ & $30$ & $512^3$ \\
GF-Strong & $35$ & $21$ & $-200$ &$9.8$ & $140$ & $512^3$ \\
\hline
\end{tabular*}

\raggedright{The mean magnetic field ($\langle B_z\rangle$) and turbulent magnetic field ($\bturb$) are in machine units (see text). The asymptotic saturation level ($\sigma_{\rm sat}$) is the total magnetic energy within the box divided by the total kinetic energy (at the beginning of the contraction phase).}
\end{table*}

In order to establish a fully-developed turbulent state, we first drive turbulence for a few crossing times. The state after this initial driving phase serves as the initial condition for our numerical experiments on the statistics of MHD turbulence in a contracting reference frame. Since we are focussing on MHD turbulence in molecular clouds, we drive turbulence to a target mass-weighted (MW) Mach number $\mach=\langle v_\mathrm{rms}/\cs \rangle_\mathrm{MW}=9$--$10$ \citep[i.e., supersonic turbulence; see][]{Larson1981,SolomonEtAl1987,OssenkopfMacLow2002,HeyerBrunt2004,HeyerEtAl2009,RomanDuvalEtAl2011,SchneiderEtAl2013} by applying a driving field $\rho{\bf F}$ as a source term in the momentum equation~(\ref{eq:mhd2}). The sound speed is chosen as $\cs=1$ in normalised units.

The turbulence driving field is constructed with a stochastic Ornstein-Uhlenbeck (OU) process \citep{EswaranPope1988,SchmidtEtAl2009,PriceFederrath2010}, implemented by \citet{FederrathDuvalKlessenSchmidtMacLow2010} and available in the public version of the FLASH code. The OU process creates a spatial and temporal driving pattern that varies smoothly in space and time with an auto-correlation timescale equal to the turbulent turnover time (also called turbulent box-crossing time), $\tturb=L/(2\mach\cs)=0.05$ for $\mach=10$ on the largest scales ($L/2$) in our periodic simulation domain of side length $L=1$ (normalised units). The driving field ${\bf F}$ is constructed in Fourier space such that most power is injected at the smallest wave numbers, $1<\left|\mathbf{k}\right|L/2\pi<3$. The peak of energy injection is on scale $L/2$, i.e., $k=2$, and falls off as a parabola towards smaller and higher wave numbers, such that the driving power is identically zero at $k=1$ and $k=3$, as in our previous studies of driven turbulence \citep[e.g.,][]{FederrathDuvalKlessenSchmidtMacLow2010,Federrath2013,Federrath2016jpp}. This procedure confines the effect of the driving to a narrow wave number range and allows the turbulence to develop self-consistently on smaller scales ($k\geq3$).

In constructing the driving field, we apply a Helmholtz decomposition in Fourier space, in order to separate the driving field into its solenoidal and compressive parts. This allows us to construct a solenoidal (divergence-free) driving field ($\nabla\cdot\mathbf{F}=0$) or a compressive (curl-free) driving field ($\nabla\times\mathbf{F}=0$). The influence of different driving on the statistics of turbulence, the amplification of magnetic fields, and on the star formation rate has been determined in \citet{FederrathKlessenSchmidt2008,FederrathKlessenSchmidt2009,FederrathDuvalKlessenSchmidtMacLow2010,federrath11}, \citet{FederrathKlessen2012,FederrathKlessen2013}, \citet{Federrath2013,Federrath2016jpp}, and \citet{FederrathEtAl2016,FederrathEtAl2017iaus}. For simplicity and since here we simply want to seed a fully-developed initial turbulent state before starting the contraction, we chose to use purely solenoidal (divergence-free) driving.

\subsection{Initial conditions and list of simulations}

We start from gas with uniform density $\rho_0=1$ (normalised units) at rest and drive turbulence for $t_0=4\,\tturb=0.2$ in a fixed (non-contracting) reference frame ($a=1$), which establishes fully-developed turbulence. After this, we begin the contraction phase, $a(t)<1$, at which point the driving is deactivated and turbulence as well as magnetic-field dynamics are solely determined by the contraction of the gas in the co-moving reference frame given by Equations~(\ref{eq:mhd1h}--\ref{eq:mhd4h}).

Table~\ref{tab:sims} provides a list of all the simulations performed. We distinguish between three main cases: two purely hydrodynamical (HD) runs, four MHD runs without magnetic guide field (noGF), and four MHD runs that include a constant guide field (GF) $\langle B_z\rangle$ in the $z$-direction of the simulation domain; for each of the MHD cases we consider multiple field strengths in order to determine the sensitivity of the results to this parameter. The simulations without guide field use an initial turbulent field generated with a flat power spectrum in the range $k/(2\pi)=2$--$20$, which produces initial turbulent fields (after the driving phase) of $\bturb=9.2$, $24$, and $21$, respectively (Table~\ref{tab:sims}, middle section). The simulations with guide field were initialised with $\langle B_z\rangle = 0.35$, $3.5$, and $35$, respectively, giving rise to initial turbulent fields (after the driving phase) of $\bturb=5.5$, $18$, and $21$, respectively (Table \ref{tab:sims}, bottom section). We note that the field strength is in normalised units, so the Alfv\'en speed $\va=B/(4\pi\rho_0)^{1/2}$ in normalised units. This means that $B \sim3.5$ corresponds to an Alfv\'en speed of one and $B\sim 35$ to an Alfv\'en speed of 10, comparable to the turbulent velocity dispersion. Thus we can think of our three guide field cases as representing three regimes of plasma $\beta$ and Alfv\'en Mach number $\mathcal{M}_A$ (computed with respect to the guide field): GF-Weak has $\beta \gg 1$, $\mathcal{M}_A \ll 1$, GF-Medium has $\beta \sim 1$, $\mathcal{M}_A \ll 1$, and GF-Strong has $\beta \ll 1$, $\mathcal{M}_A \sim 1$.

All our simulations use the same resolution of $N_\mathrm{res}^3=512^3$ grid cells (except MHD-noGF-Strong-LR with $N_\mathrm{res}=256$, used to investigate numerical convergence in Appendix~\se{convergence}).

Finally, our simulations (Tab.~\ref{tab:sims}) use the same time evolution for the scale factor $a(t)=\exp{[H(t-t_0)]}$ for $t\ge t_0$ with $H=-(\tturb/10)^{-1}=-200$, i.e., fast contraction on a time scale ten times shorter than the initial turbulent crossing time. However, we also run an HD-H1, MHD-noGF-Strong-H1 and MHD-GF-Medium-H1 simulation with $H=-1$, in order to demonstrate that our main conclusions do not depend on the choice of $H$ (see Appendix~\se{H1}). \citet{robertson12} discussed three different cases for the contraction law in the pure HD limit, while here we are primarily interested in the case of fast contraction (compression), focussing on the effect of the magnetic field (MHD runs).

% SECTION: Results
\section{Simulation Results}
%\section{Simulations of contraction-driven MHD turbulence}

\label{sec:results}

\subsection{Evolution of the Mach number and energy}

\begin{figure}
	\centering
	\includegraphics[width=0.475\textwidth]{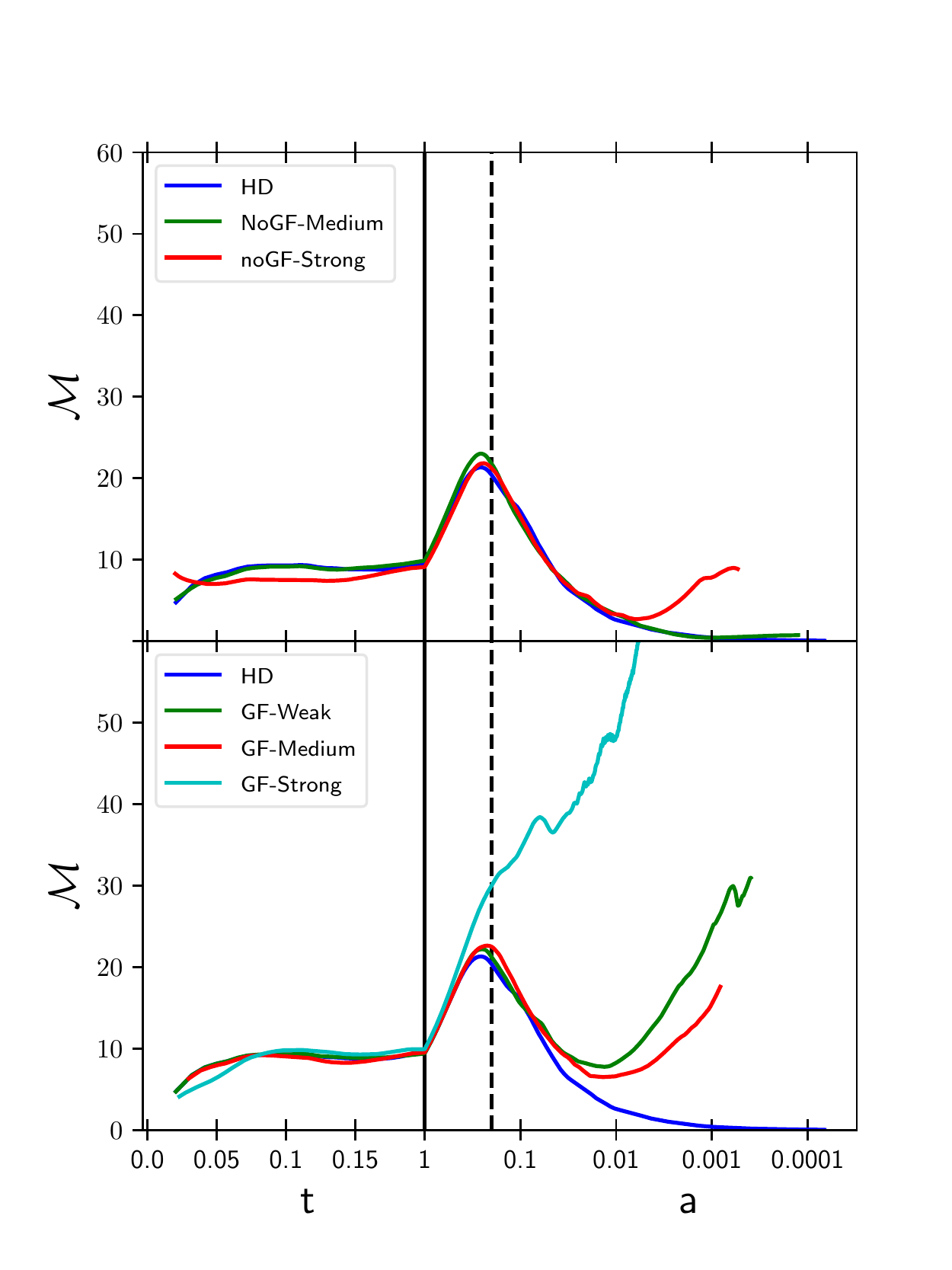}
	\caption{The mass-weighted root mean square Mach number of the turbulent gas as a function of time ($t$, to the left of the vertical solid line) and scale factor ($a$, to the right of the vertical solid line).
The vertical solid line separates between the initial driving stage, when turbulence driving is active (\S\ref{sec:driving}) and the box is static ($a=1$), and the compression stage, when the driving is disabled and the box contracts ($a(t)<1$).	
	 {\it Top panel:} simulations without magnetic guide field (``noGF'').  {\it Bottom panel:} simulations with magnetic guide field (``GF'') (see Table~\ref{tab:sims}). We show the no-magnetic field case (``HD") in both panels to guide the eye. The dashed vertical line indicates the value for which the dissipation rate equals the compression rate, following \equ{dissip}.}
	\label{fig:mach_a}
\end{figure}

\begin{figure}
	\centering
	\includegraphics[width=0.475\textwidth]{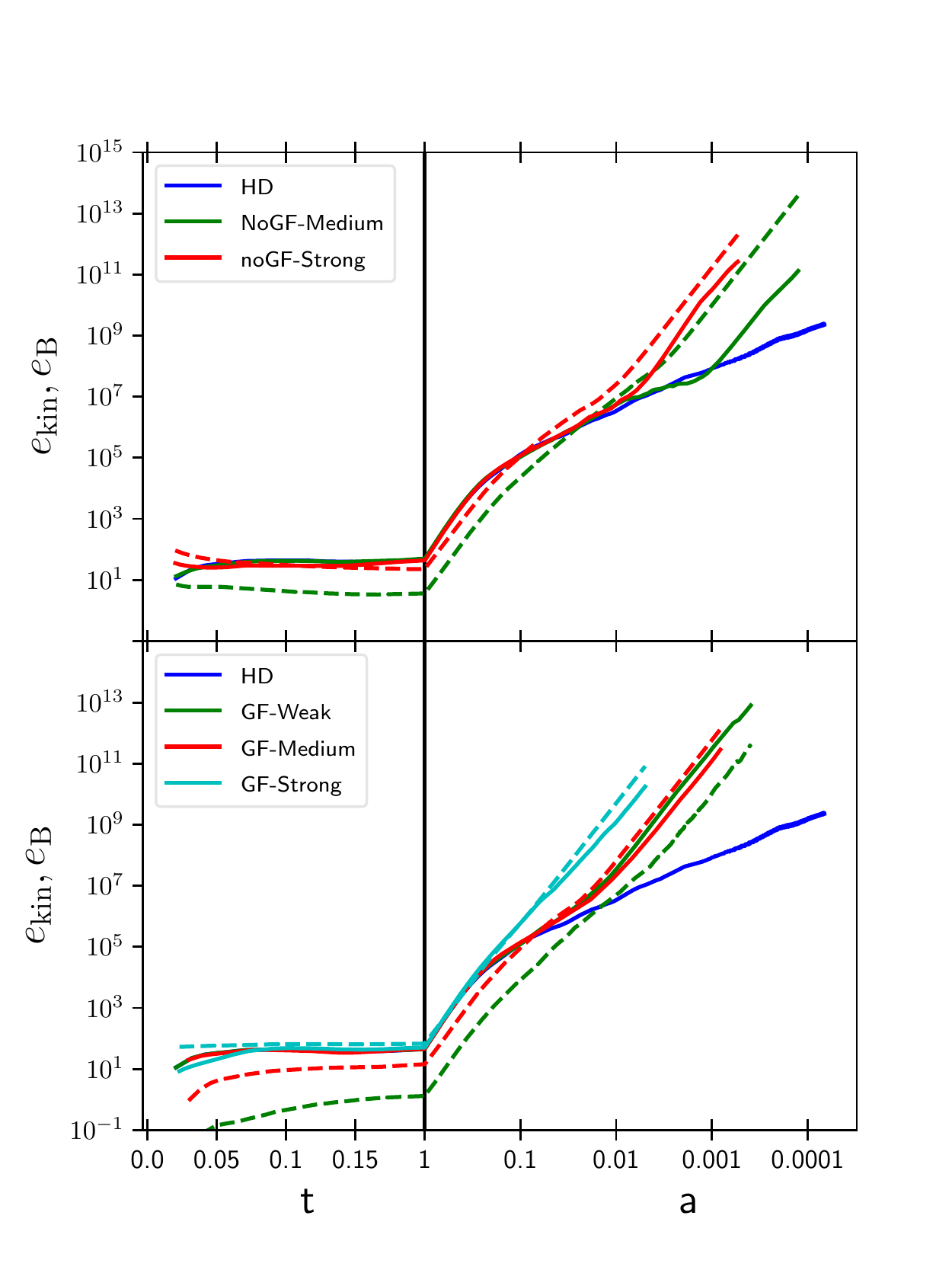}
	\caption{Same as \figref{mach_a}, but showing the kinetic (solid lines) and magnetic (dashed lines) energy per unit volume, in normalised units (see~\S\ref{sec:sims}). We list the ratio $\sigma_{\rm sat} \equiv e_{\rm B}/e_{\rm kin}$ at $a=1$ in Tab.~\ref{tab:sims}.}
	\label{fig:E_t}
\end{figure}

\figref{mach_a} and \ref{fig:E_t} show the time evolution of the rms Mach number and the kinetic and thermal energies, respectively, in all simulations. We show these quantities as a function of time during the initial driving phase, and as a function scale factor once compression begins, with the two phases separated by the solid vertical lines in the plots. Since the scale factor is exponential in time ($H=\dot{a}/a=-200$) and the plots use a logarithmic scale, position on the $x$-axis is proportional to time during both phases, albeit with different scalings. The top panels show our noGF simulations (without magnetic guide field) and the bottom panel shows our GF simulations. We show the pure hydrodynamic simulation (HD) in both panels to help guide the eye. 

First examine \figref{mach_a}. We see that, starting from the fully-developed turbulent state at $a=1$, all simulations start compression (decreasing $a$), which drives turbulence, i.e., increasing $v$ and $\mach$. The turbulence is initially supersonic with $\mach\sim9$--$10$ (Tab.~\ref{tab:sims}), and increases to a peak of $\mathcal{M}\approx 20$. However, at $a\sim0.2$, in all simulations except noGF-Strong, the evolution reverses and turbulence begins to decay. This change is also apparent in the kinetic energy density evolution shown in \figref{E_t}, which increases sharply from $a=1$ to $a\approx 0.2$, but then shows an inflection point and increases less steeply thereafter.

This qualitative change from increasing to decaying turbulence is not related to the turbulence becoming sonic or subsonic, which only happens much later. Instead, it can be explained by the change of dissipation with $a$. The dissipation timescale is proportional to the largest eddy turnover time \citep{maclow99,robertson12} and the dissipation rate becomes comparable to the compression rate when
\begin{equation}
    \eta^* \frac{v_{\rm rms}(a)}{a\lambda}=\eta \frac{v_{\rm rms}}{aL}=|H|=-\frac{\dot{a}}{a},
    \label{eq:dissip}
\end{equation}
with $v_\mathrm{rms}$ the root mean square of the velocity field, $\lambda=L/2$ the largest eddy size and $H=-200$, the compression rate in our simulations (\S\ref{sec:sims}). The coefficient $\eta^*=\eta/2$ is a dimensionless dissipation efficiency (see \se{model}) and has been calibrated  to be $\eta^* \approx 0.9$ (see \se{comparison}). We can solve \equ{dissip} numerically for $a$ using the value of $v_{\rm rms}(a)$ measured from the simulations, and the result is $a\approx 0.2$; we show the exact solution for the HD run as the vertical dashed line in \figref{mach_a}. It is evident that  this typical timescale for equality between compression and dissipation successfully predicts the onset of efficient dissipation and roughly coincides with the beginning of the decaying stage of turbulence. 

From $a=1$ to $a\approx 0.2$, the magnetic field has only minor effects on the evolution in all runs  except GF-Strong. Compared to the HD case, in the noGF models the magnetic field stores additional energy which replenishes some of the kinetic energy that is dissipated. This slightly delays the onset of the decaying stage, and allows higher maximum velocities or Mach numbers by about $10$--$20\%$, but this is clearly a modest effect. However, at later times the MHD and HD runs show profound differences. In all the MHD runs the Mach number (\figref{mach_a}) eventually stops decreasing and begins to increase again. Corresponding to this, the slope of the kinetic energy versus $a$ curve (\figref{E_t}) steepens again. 

The value of $a$ at which the switch from decaying to increasing turbulence happens appears to depend both on whether there is a guide field, and on the saturation level of the turbulent dynamo, as parameterized by $\sigma_{\rm sat} \equiv e_{\rm B}/e_{\rm kin}$, at the onset of compression; we report this quantity in Tab.~\ref{tab:sims}. The noGF-medium run has $\sigma_{\rm sat} = 8\%$, and does not switch from decaying to increasing until $a\approx 0.001$, while the noGF-Strong ($\sigma_{\rm sat} = 56\%$), GF-Weak ($\sigma_{\rm sat} = 3\%$), and GF-Medium ($\sigma_{\rm sat} = 30\%$) all reverse at $a\approx 0.01$. The GF-Strong case ($\sigma_{\rm sat} = 130\%$) never goes through a decaying phase at all, and instead has a Mach number that increases almost monotonically. In call cases, the difference between the MHD and HD cases is large and growing with time. Even the noGF-Medium case, with $\sigma_{\rm sat} = 8\%$, has $\sim 10$ times as much kinetic energy as the pure HD case by $a=0.001$. The GF-Strong case has 10 times the kinetic energy of the HD case even at $a\approx 0.1$, and by $a = 0.01$ this gap has grown to more than two orders of magnitude.

\subsection{Dissipationless flows}
\label{ssec:dissipationless}

\subsubsection{The transition to dissipationless flow}

\begin{figure}
	\centering
	\includegraphics[width=0.475\textwidth]{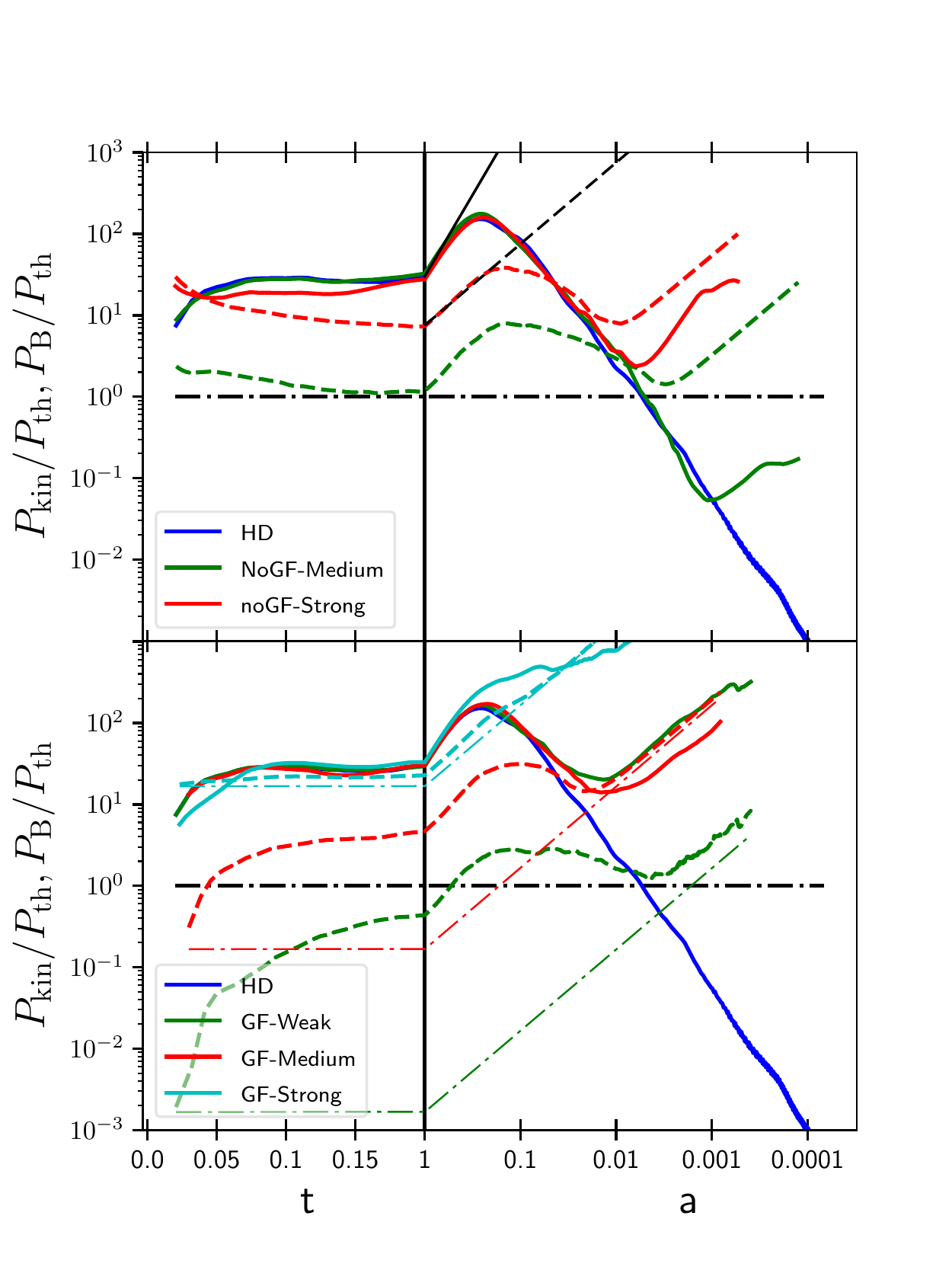}
	\caption{The ratio of kinetic to thermal pressure (solid lines), and magnetic to thermal pressure (dashed lines) for the noGF simulations (\top{}) and GF simulations ({\it bottom panel} -- see \cref{tab:sims} for simulation details). The dot-dashed horizontal line marks the sonic ratio, where the pressure equals the thermal pressure (which is equivalent to Mach number $\mach=\sqrt3$). The slanted black solid and dashed lines in the {\top} show the expected adiabatic compression slope of the kinetic and magnetic pressure ratios (see text). The dot-dashed lines in the \bot{} show the magnetic pressure that results from the volume-averaged $z$-component of the magnetic field. 
}
	\label{fig:beta_mach}
\end{figure}

\begin{figure*}
\centering
\subfigure[HD, $a=1$.]{
\includegraphics[scale=0.09]{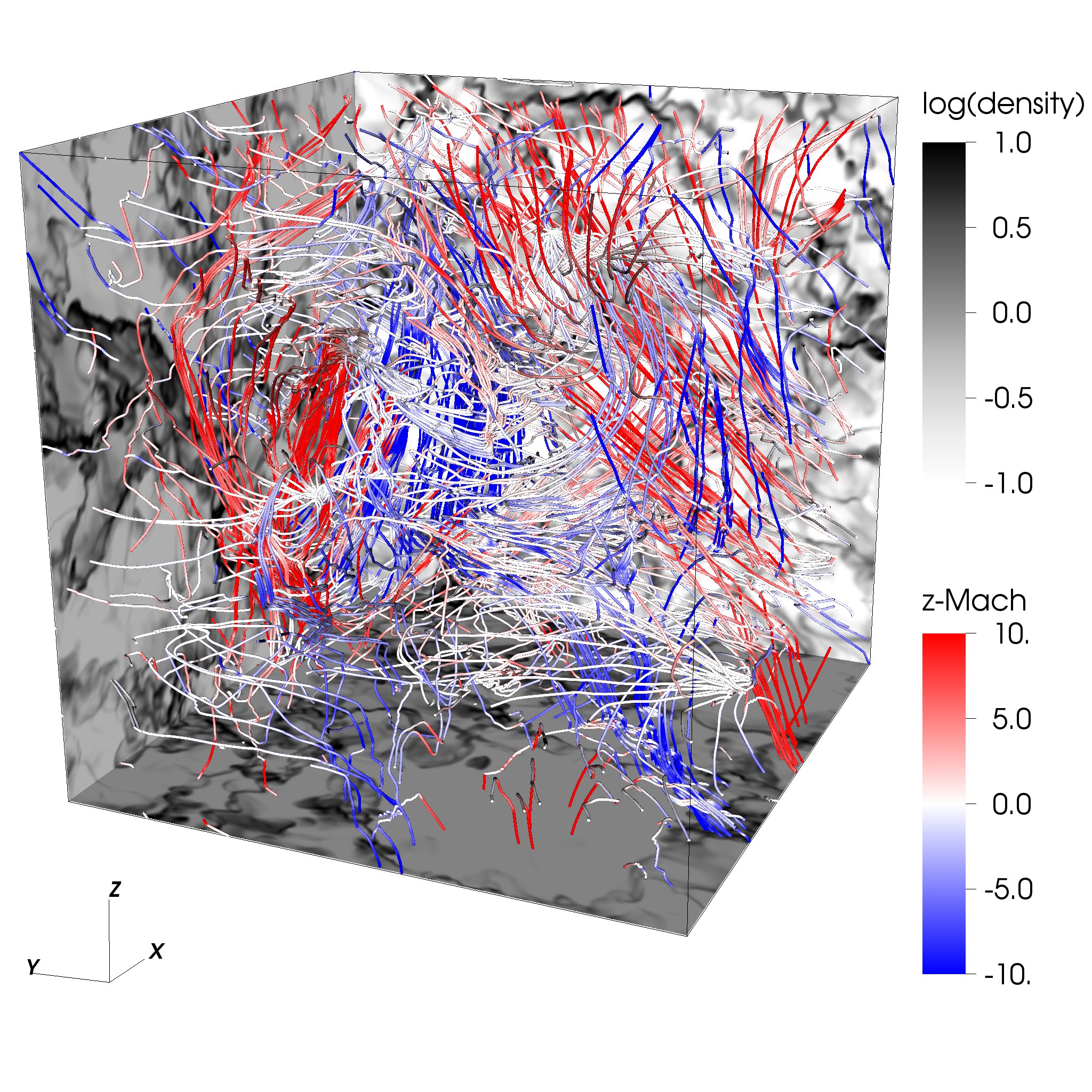}}
\subfigure[HD, $a=10^{-3}$.]{
\includegraphics[scale=0.09]{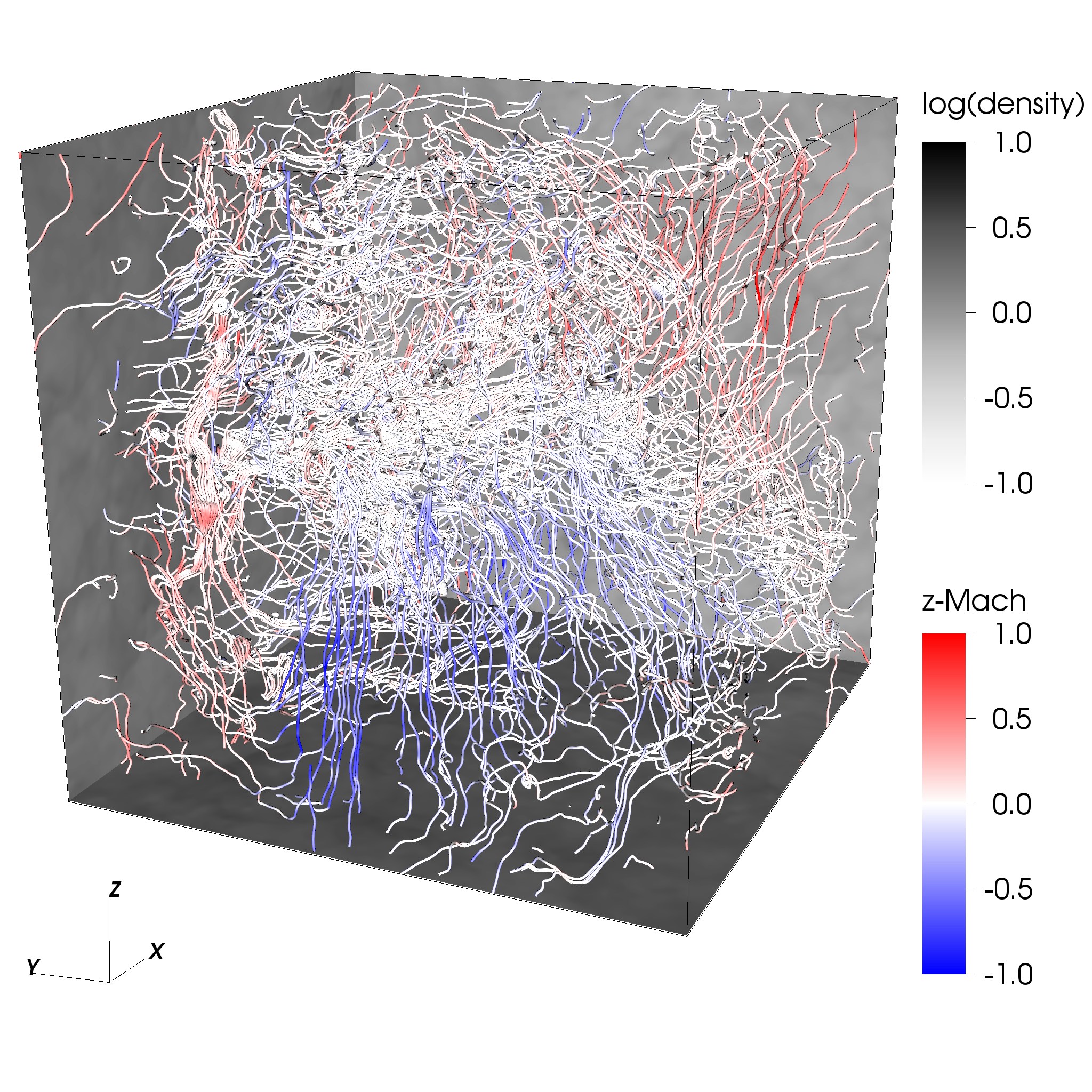}}
        \vskip\baselineskip
\subfigure[MHD-noGF-Strong, $a=1$.]{
\includegraphics[scale=0.09]{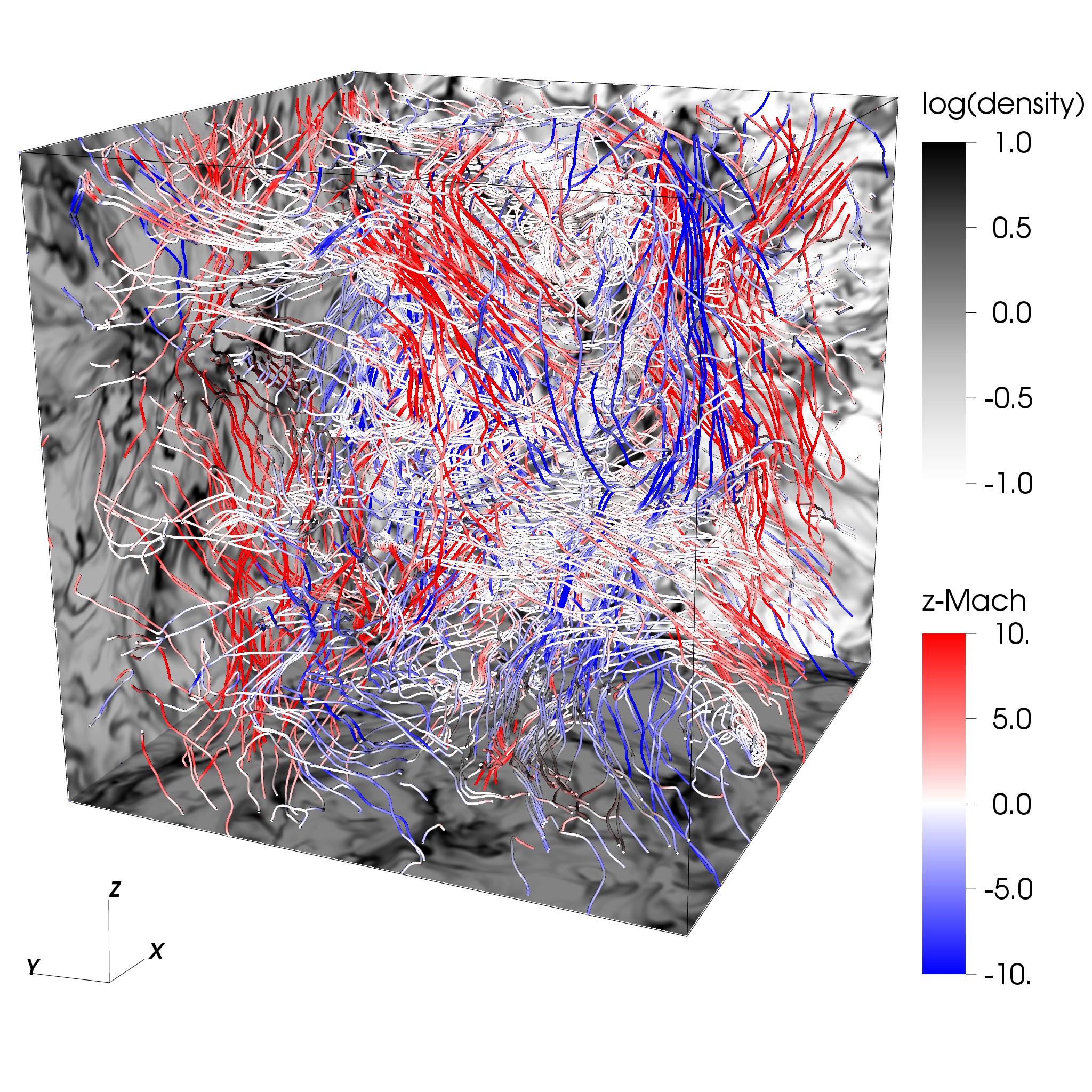}}
\subfigure[MHD-noGF-Strong, $a=10^{-3}$.]{
\includegraphics[scale=0.09]{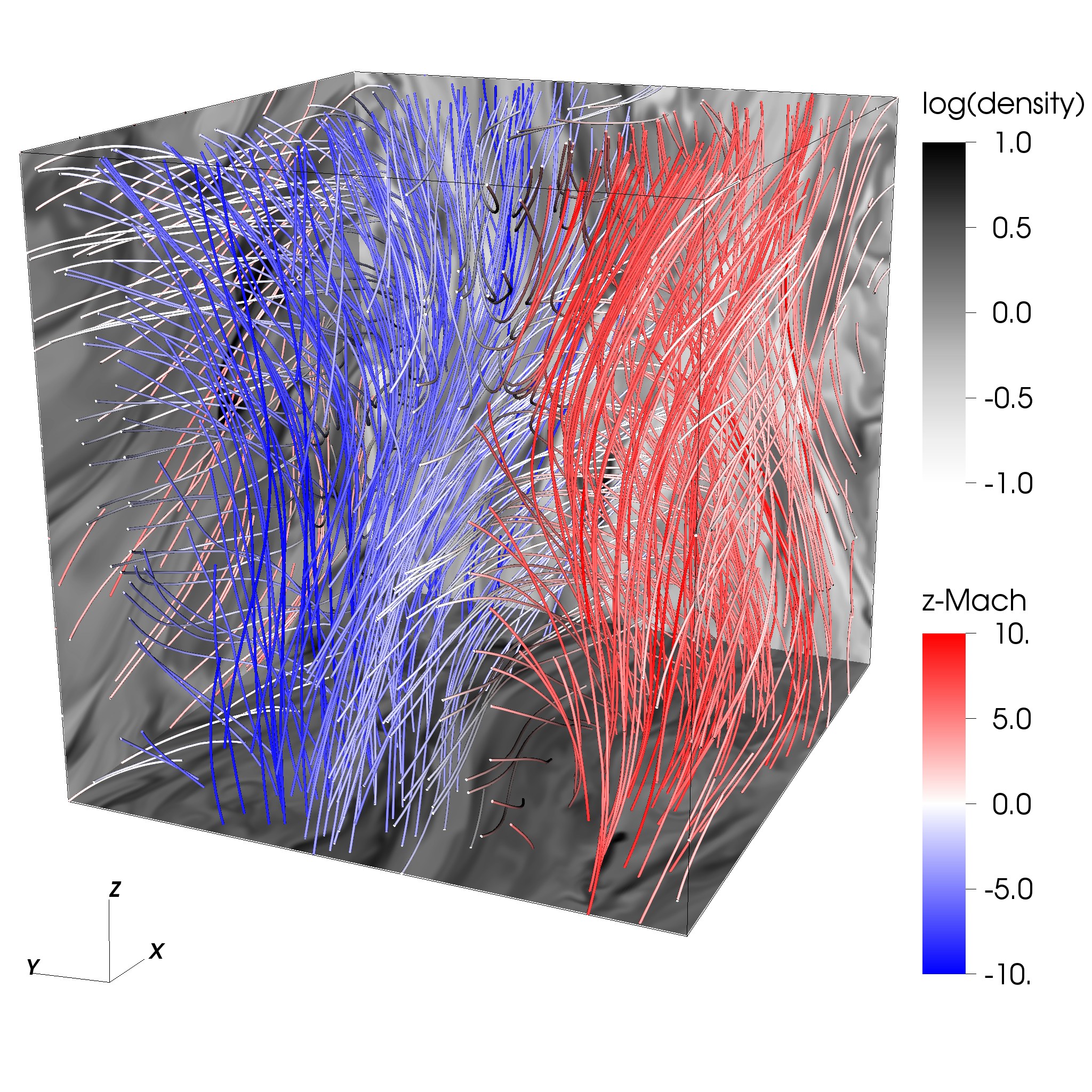}}
        \vskip\baselineskip
\subfigure[MHD-GF-Medium, $a=1$.]{
\includegraphics[scale=0.09]{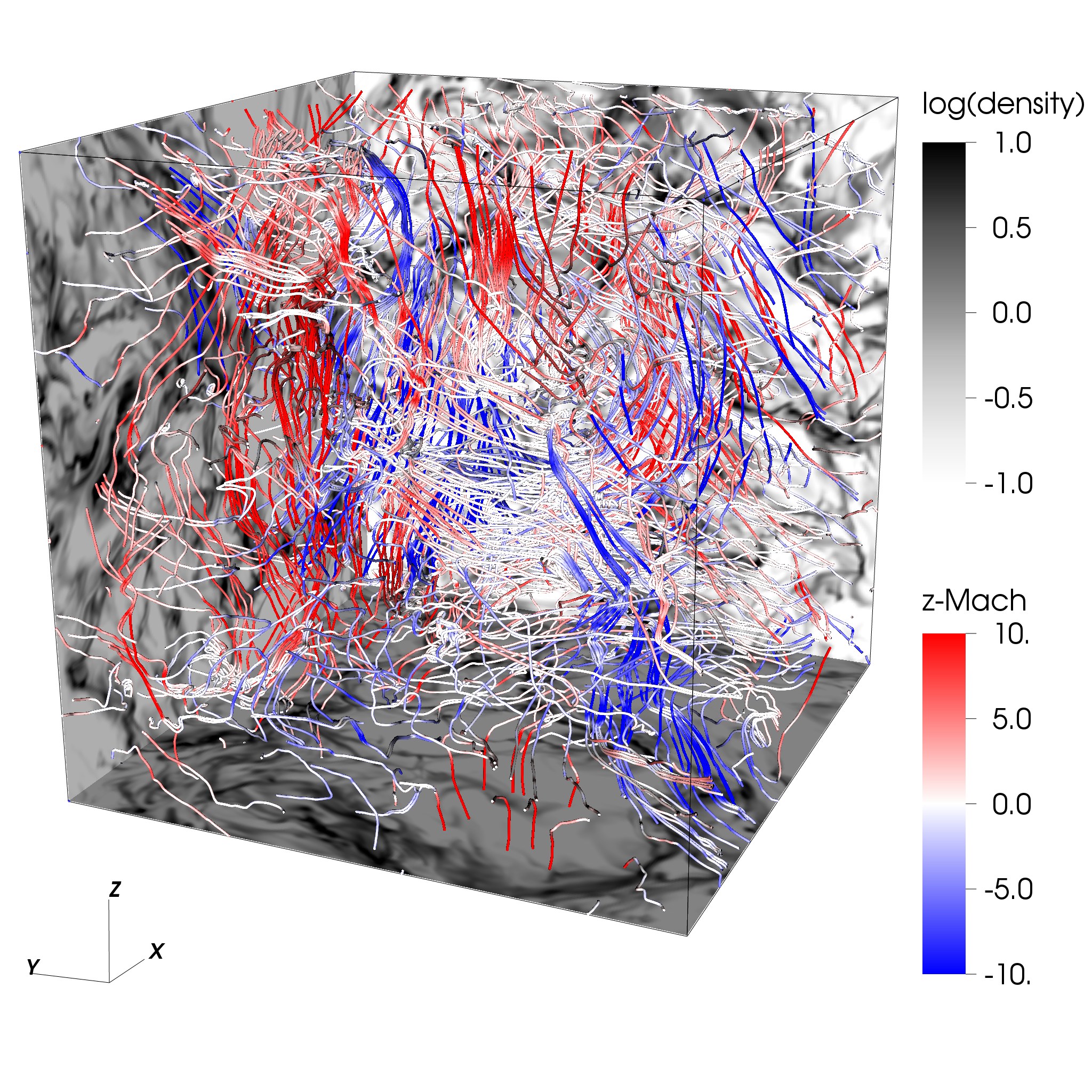}}
\subfigure[MHD-GF-Medium, $a=10^{-3}$.]{
\includegraphics[scale=0.09]{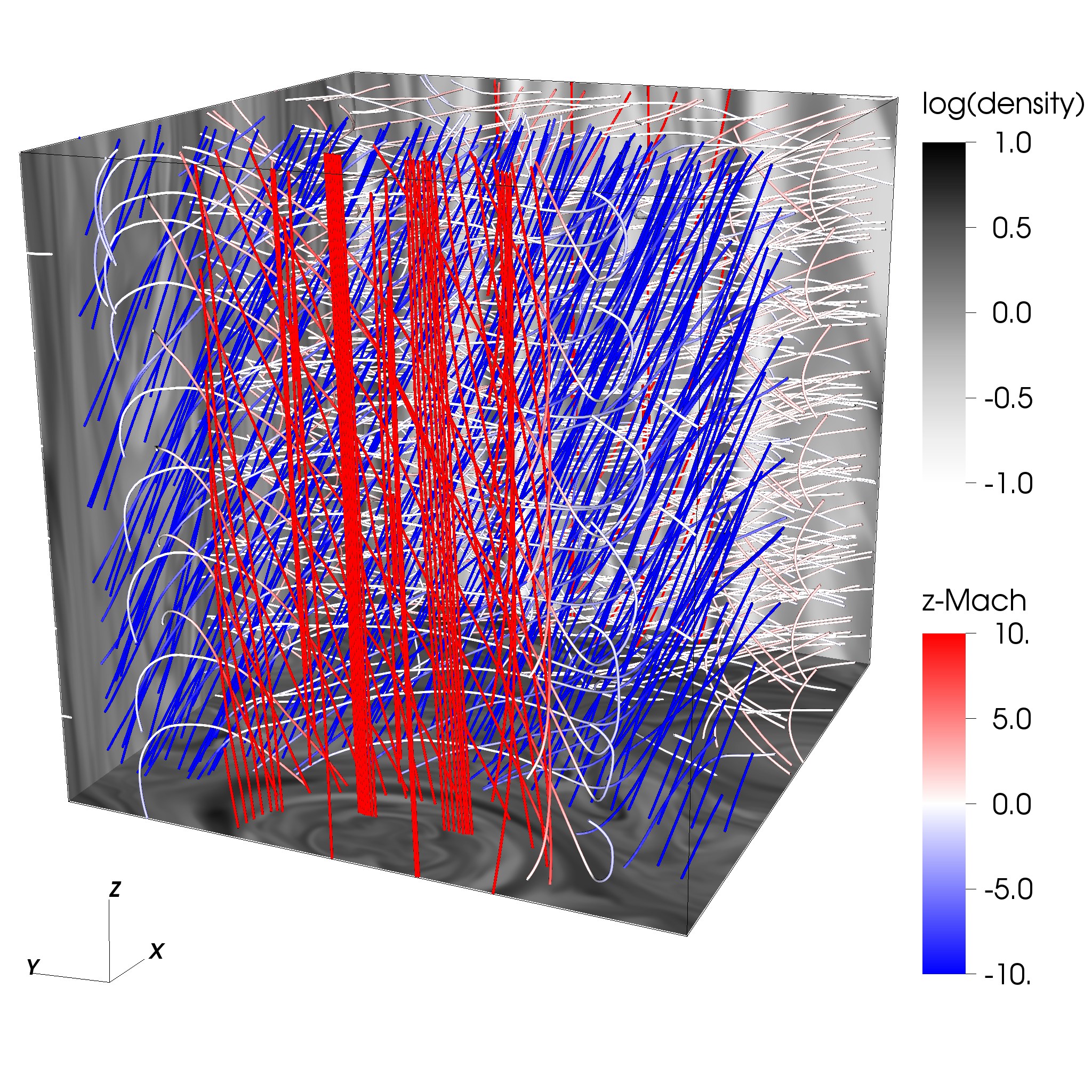}}
\caption{\label{fig:morphology} Flow morphology for pure HD ({\it top row}), noGF-Strong ({\it middle row}) and GF-Medium ({\it bottom row}) at beginning of contraction ({\it left column}) and at $a=10^{-3}$ ({\it right column}). Lines show Mach numbers of the {\it z}-component of velocity along streamlines for the flow. Background colourmaps show co-moving density slices along the principle directions.} 
\end{figure*}

\begin{figure}
	\centering
	\includegraphics[width=0.475\textwidth]{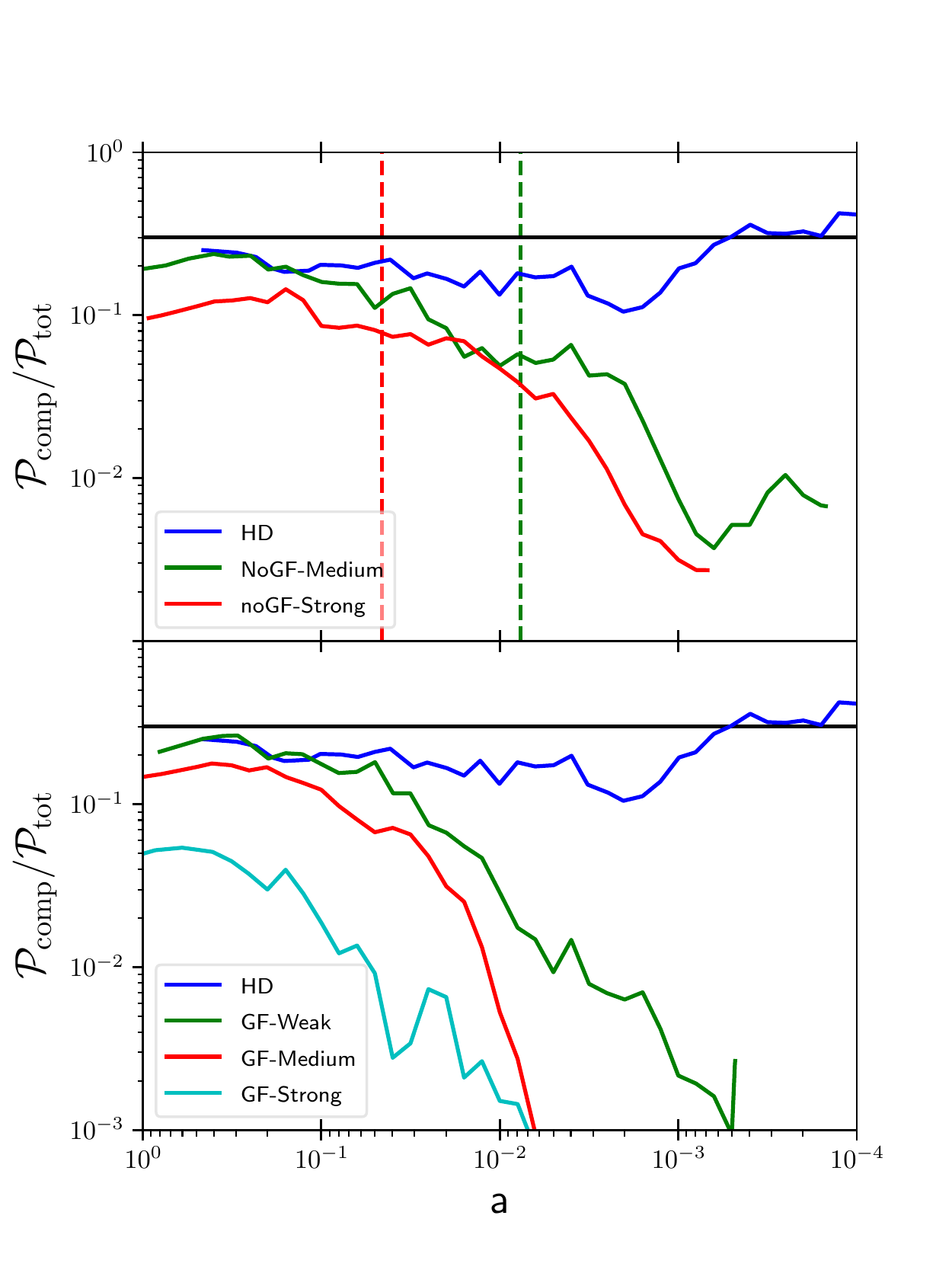}
	\caption{The compressive ratio, $\langle v_\mathrm{c}^2\rangle / (\langle v_\mathrm{s}^2\rangle\!+\!\langle v_\mathrm{c}^2\rangle)$, of the flow, as a function of the scale factor, for the HD and noGF (\top) and GF (\bot) simulations. For supersonic turbulence, this value is expected to be between $\sim20$--$60\%$, depending on the driving mode of the turbulence \citep{federrath11}. A lower value indicates that the flow is dominated by incompressible modes, which is clearly seen for the runs that include magnetic fields. The vertical lines show the scale factor for which the magnetic pressure dominates over the kinetic pressure (\figref{beta_mach}). For scale factors below this transition, the compressive modes decay and the system is dominated by incompressible modes. Compressible modes only contribute $<1\%$ of the energy at late times for the MHD runs, while the HD run maintains high compressive ratios of $\sim10$--$40\%$.}
	\label{fig:Elgt}
\end{figure}

Having seen that the presence of a magnetic field causes a major change in the behavior of compressive turbulence, we now investigate in more detail the origin of this behavior. We shall show that this change the result of a shift in the flow pattern to one that is nearly dissipationless. As a first step in this direction, we note that the switch from decaying to increasing Mach number is associated the ratio of magnetic to kinetic pressure. In our dimensionless units, the volume-averaged thermal pressure is $1/V$, where $V$ is the box volume, and we define the volume-averaged kinetic and magnetic pressures by
\begin{eqnarray}
\label{eq:pkin_sim}
P_{\rm kin} & = & \frac{1}{V}\int \frac{1}{2} \rho v^2 \, dV \\
\label{eq:pb_sim}
P_{\rm B} & = & \frac{1}{V} \int \frac{1}{3} \left(\frac{B^2}{8\pi}\right) \, dV.
\end{eqnarray}
Note that the factor of $1/3$ in the definition of $P_{\rm B}$ might at first seem surprising, but we shall see the justification for it in \S\ref{sec:model}. 

We plot the time evolution of $P_{\rm kin}$ and $P_{\rm B}$ in all our runs in \figref{beta_mach}. As in \cref{fig:mach_a,fig:E_t}, the $x$-axis is separated (by the vertical solid line) into the initial driving stage on the left, and the contraction phase on the right. At the beginning of compression, dissipation is comparatively unimportant because the compression timescale is small compared to the eddy turnover timescale. Thus the flow is nearly dissipationless. We show below that, for adiabatic contraction, kinetic pressure acts as a gas with $\gamma = 5/3$ and turbulent magnetic pressure acts as a gas with $\gamma = 4/3$, and we expect these pressures to scale as
\begin{equation}
\frac{P}{P_{\rm th}}\propto \rho^{\gamma-1}\propto a^{-3(\gamma-1)}.
\end{equation}
Thus we expect the kinetic to thermal ratio to scale as $a^{-2}$, and the magnetic to thermal ratio to scale as $a^{-1}$ for the case without a guide field. With a guide field, flux conservation requires that the mean magnetic field rise as $B_{\rm mean} \propto a^{-2}$, and thus the scaling is the same, though for a somewhat different reason. We show lines with slopes of $-1$ and $-2$ in \figref{beta_mach}, and they are indeed good descriptions of the slope at early times.

Unsurprisingly, the kinetic and magnetic pressures begin to drop when the dissipation rate becomes comparable to the compression rate. However, the kinetic term drops more steeply than the magnetic term. This effect is due to the fact that the only true dissipation channel in the system is via the kinetic term, and that the dissipation of the magnetic component is bottlenecked by the rate at which the now over-magnetized gas can transfer energy back into the kinetic component.

As a result of the difference in the dissipation rates, the magnetic pressure ultimately exceeds the kinetic pressure in all cases except GF-Weak. In all the other cases, the transition from decreasing to increasing Mach number occurs almost exactly when this crossover happens, although if one closely compares GF-Medium to noGF-Medium, it is clear that at equal field strength the transition occurs earlier, in terms of both $a$ and in terms of ratio of $P_{\rm kin}$ to $P_{\rm B}$, in the presence of a guide field. Whether the flow is subsonic or supersonic appears to make little difference to the transition, consistent with the findings of \citet{maclow99} that the dissipation rate is not greatly affected by whether the flow is subsonic or supersonic.

\subsubsection{The nature of the dissipationless flow}

When the magnetic pressure begins to dominate, or even earlier in the presence of a net magnetic flux, the flow re-arranges itself into a fundamentally different topology, characterized by a much lower rate of dissipation. We illustrate this topology in \figref{morphology}, which shows density field maps along the major three axes, and velocity streamlines colour-coded by their $z$-component of the Mach number. The {\it left column} presents the initial state ($a=1$) and the {\it right column} a highly compressed stage for which the flow has had time to settle into a self consistent non-driven mode ($a=10^{-3}$). The runs presented here are the HD run ({\it top}), noGF-Strong ({\it middle}) and GF-Medium ({\it bottom} panels), but the other MHD runs are qualitatively the same as the two shown in the figure.

Both the MHD runs exhibit a behaviour such that after compression has taken place, the flow settles into two main sheets sliding across each other at supersonic velocities. Since there is no preferred direction in the noGF run, and in the x-y plane of the GF run, in the simulation setup, the division into the domains is arbitrary (for this specific simulation it roughly coincides with the $x$-axis. It is clear that this flow, which has naturally developed from standard turbulence, is highly non-random, and that the expected dissipation of these flows is greatly reduced compared to the standard flow of the HD run or the initial state.

We can illustrate the reduces dissipation more directly by examining the ratio of flow power in compressible modes to the total power in all modes, which we refer to as the compressive ratio. We show the time evolution of this quantity for all runs in  \figref{Elgt}. We compute the energy in solenoidal ($\langle v_\mathrm{s}^2\rangle$) and compressible ($\langle v_\mathrm{c}^2\rangle$) modes by performing a Helmholtz decomposition of the velocity field \citep{FederrathDuvalKlessenSchmidtMacLow2010,federrath11,PanEtAl2016,JinEtAl2017}. The ratio of $\langle v_\mathrm{c}^2\rangle / (\langle v_\mathrm{s}^2\rangle\!+\!\langle v_\mathrm{c}^2\rangle) \sim 0.2$--$0.6$ for supersonic turbulence depends on the driving mode \citep{federrath11}. In the absence of magnetic fields, the flow remains in this range of values even after driving ceases, during the compressive phase (the HD case). However, \figtop{Elgt} demonstrates that when the magnetic field begins to dominate ($a\lesssim 0.01$ for noGF-Medium, and $a\lesssim 0.03$ for noGF-Strong; see also \figref{beta_mach}, \top{}), the compressive ratio drops rapidly. These values are marked by the vertical dashed lines in \figtop{Elgt}.

The GF-Strong run is particularly noteworthy in that it has a compressive ratio $\lesssim 10\%$ even before the onset of compression, simply as a result of the strong magnetic field that prevents flows across field lines. As a result, it never experiences significant dissipation, and never goes through a phase when the turbulence decays.

% SECTION
\section{Theoretical framework for compressible MHD turbulence} \label{sec:model}

Having seen that magnetic fields lead to novel and initially-unexpected effects in MHD turbulence, we now seek to construct a theoretical model that we can use to interpret the results. Our basic approach will be to think of the region we are simulating as a small portion of a much larger cloud. We will then coarse-grain the MHD equations over the scale of our box, allowing us to write down an effective pressure in the box. We will use the results of our numerical experiments, together with some basic physical arguments, to provide an effective equation of state to describe this pressure and its evolution, so that we can interpret our numerical results in thermodynamic terms.

\subsection{Coarse-grained pressures}

We begin by following the usual method of constructing a set of coarse-grained equations \citep[e.g.,][]{Germano92a, kuncic04, Schmidt06a, Schmidt11a}. We define a spatial filter $F_\Delta(\mathbf{x})$ with characteristic scale $\Delta$ with which we can convolve all the fluid variables. For any field $\phi(\mathbf{x})$, we define
\begin{eqnarray}
\overline{\phi} & \equiv & \int \phi(\mathbf{x}') F_\Delta(\mathbf{x}-\mathbf{x}') \,d\mathbf{x}' \\
\phi' & \equiv & \phi - \overline{\phi} \\
\tilde{\phi} & \equiv & \frac{\overline{\rho\phi}}{\overline{\rho}}.
\end{eqnarray}
Here $\overline{\phi}$ is the filtered variable, obtained by convolving $\phi$ with the filter, and $\phi'$ is the fluctuating part that remains after the filtered part has been removed.

Convolving the MHD equation of momentum conservation, \equ{mhd2}, with the filter $F_\Delta(\mathbf{x})$ gives
\begin{eqnarray}
0 & = &
\frac{\partial}{\partial t}\left(\overline{\rho}\tilde{\mathbf{v}}\right) +
\nabla\cdot\left(\overline{\rho\mathbf{v}\otimes\mathbf{v}} - 
\frac{1}{4\pi}\overline{\mathbf{B}\otimes\mathbf{B}} + \frac{1}{8\pi}\overline{B^2}\mathbf{I}\right)
\nonumber \\
& & {} + \nabla \overline{P_{\rm th}},
\end{eqnarray}
where $\mathbf{I}$ is the identity tensor. Per the usual approach, we now write the averages over correlated terms as differences of the filtered quantities and the sub-filter-scale (SFS) quantities,
\begin{eqnarray}
0 & = &
\frac{\partial}{\partial t}\left(\overline{\rho}\tilde{\mathbf{v}}\right) +
\nabla\cdot\left(\overline{\rho}\tilde{\mathbf{v}}\otimes\tilde{\mathbf{v}} - 
\frac{1}{4\pi}\overline{\mathbf{B}}\otimes\overline{\mathbf{B}} + \frac{1}{8\pi}\overline{B}^2\mathbf{I}\right)
\nonumber \\
& & {} + \nabla \overline{P_{\rm th}} - \nabla\cdot\left(\taub_{\rm R,SFS} + \taub_{\rm M,SFS}\right),
\end{eqnarray}
where
\begin{eqnarray}
\taub_{\rm R,SFS} & = & \overline{\rho}\tilde{\mathbf{v}}\otimes\tilde{\mathbf{v}} - \overline{\rho\mathbf{v}\otimes\mathbf{v}} \\
\taub_{\rm M,SFS} & = & -\frac{1}{4\pi}\overline{\mathbf{B}}\otimes\overline{\mathbf{B}} +
\frac{1}{8\pi} \overline{B}^2 
\nonumber \\
& & {} +  \frac{1}{4\pi}\overline{\mathbf{B}\otimes\mathbf{B}} -
\frac{1}{8\pi} \overline{B^2}
\end{eqnarray}
are the Reynolds stress and Maxwell stress exerted by the SFS components of the fluid velocity and magnetic field, respectively.

As standard for the microphysical stress tensor, we decompose the SFS stresses $\taub_{\rm R,SFS}$ and $\taub_{\rm M,SFS}$ into on- and off-diagonal components, and identify the former as effective pressures. That is, we define the effective kinetic and magnetic pressures by
\begin{eqnarray}
P_{\rm kin} & = & -\frac{1}{3} \mbox{tr} \,\taub_{\rm R,SFS} \\
\taub_{\rm R,SFS} & = & -P_{\rm kin} \mathbf{I} + \bm{\pi}_{\rm R,SFS} \\
P_{\rm B} & = & -\frac{1}{3} \mbox{tr} \,\taub_{\rm M,SFS} \\
\taub_{\rm M,SFS} & = & -P_{\rm B} \mathbf{I} + \bm{\pi}_{\rm M,SFS}.
\end{eqnarray}
We use the notation $P_{\rm B}$ for the effective magnetic pressure to distinguish it from $P_{\rm mag}$, the true, microphysical magnetic pressure, since we shall see below that they are somewhat different. For homogenous, isotropic turbulence the tensors $\bm{\pi}_{\rm R,SFS}$ and $\bm{\pi}_{\rm M,SFS}$ have zero on their diagonals. In the presence of a large-scale guide field where isotropy is broken, this is not necessarily the case, and in principle the on-diagonal components of $\bm{\pi}_{\rm R,SFS}$ and $\bm{\pi}_{\rm B,SFS}$ can be as large as $P_{\rm kin}$ and $P_{\rm B}$. However, since we are only after a heuristic model, we will ignore this complication. With these definitions, the filtered momentum equation reads
\begin{eqnarray}
0 & = &
\frac{\partial}{\partial t}\left(\overline{\rho}\tilde{\mathbf{v}}\right) +
\nabla\cdot\left(\overline{\rho}\tilde{\mathbf{v}}\otimes\tilde{\mathbf{v}} - 
\frac{1}{4\pi}\overline{\mathbf{B}}\otimes\overline{\mathbf{B}} + \frac{1}{8\pi}\overline{B}^2\mathbf{I}\right)
\nonumber \\
& & {} + \nabla \left(\overline{P_{\rm th}}+P_{\rm kin} + P_{\rm B}\right) 
\nonumber \\
& & {} - \nabla\cdot\left(\bm{\pi}_{\rm R,SFS} + \bm{\pi}_{\rm M,SFS}\right).
\end{eqnarray}

The final step in defining the coarse-grained pressures via an equation of state is to relate the pressures as we have defined them to the energy content of the gas. The SFS kinetic and magnetic energies per unit volume are simply the differences between the true energies per unit volume and their analogs defined using the filtered quantities, i.e.,
\begin{eqnarray}
e_{\rm kin,SFS} & = & \frac{1}{2}\overline{\rho v^2} - \frac{1}{2} \overline{\rho}\tilde{\mathbf{v}}^2 \\
e_{\rm B,SFS} & = & \frac{\overline{B^2}}{8\pi} - \frac{\overline{B}^2}{8\pi}.
\end{eqnarray}
From the definitions of $\taub_{\rm R,SFS}$, $\taub_{\rm M,SFS}$, $P_{\rm kin}$, and $P_{\rm B}$, it is immediately clear that we have
\begin{eqnarray}
P_{\rm kin} & = & \frac{2}{3} e_{\rm kin,SFS} \\
P_{\rm B} & = & \frac{1}{3} e_{\rm B,SFS},
\end{eqnarray}
i.e., the kinetic pressure is simply $2/3$ of the sub-filter-scale kinetic energy density, and the magnetic pressure is $1/3$ the scale of the sub-filter-scale magnetic energy density. Note that this relationship between pressure and energy density is different than the ones that obtain between the microscopic pressures and energy densities, for which $P_{\rm th} = (\gamma-1) e_{\rm th}$ and $P_{\rm mag} = e_B = B^2/8\pi$. This difference is the reason for the factor of $1/3$ we introduce into $P_B$ as computed in eq.~\ref{eq:pb_sim}. Interpreted in terms of an adiabatic index $\gamma$, we see that SFS kinetic pressure acts like a fluid with a $\gamma=5/3$ equation of state, while SFS magnetic pressure acts like a fluid with a $\gamma=4/3$ equation of state.

If the thermal pressure also obeys an equation of state
\begin{equation}
    P=\left(\gamma-1\right)\rho \epsilon,
    \label{eq:EoS}
\end{equation}
where $\epsilon$ is the energy per unit mass, then we can write the total coarse-grained pressure as
\begin{eqnarray}
P_{\rm tot} & = & \overline{P_{\rm th}} + P_{\rm kin} + P_{\rm B}
    \label{eq:ptot}
\\
& = & 
\overline{\rho} \left[(\gamma-1)\epsilon_{\rm th} +
\frac{2}{3} \epsilon_{\rm kin,SFS} + \frac{1}{3} \epsilon_{\rm B,SFS}\right],
\end{eqnarray}
where the various $\epsilon$ terms are the thermal, SFS kinetic, or SFS magnetic energy per unit mass.

\subsection{An effective EoS for supersonic magnetized gas}
\label{sec:gamma}

We now wish to model the reaction of the full system of isothermal, turbulent and magnetized gas to compression, taking into account dissipation terms and interactions between the various components. Since these additional energy transfers are time dependent, they cannot be modelled as a {\it proper} EoS, that is only a function of the thermodynamic state. Instead, we model this behaviour as an {\it effective} EoS which also depends on the thermodynamic trajectory of a parcel of gas. A similar approach has been successfully implemented for stability analysis of gravitationally collapsing haloes, filaments and sheets in a cosmological context in \citet{bd03}, \citet{db06}, and \citet{birnboim16}. For the sake of notational simplicity, we shall from this point forward drop the overlines and the SFS notation, and we will understand that, unless otherwise stated, all quantities except energies are unit mass are filtered quantities, while specific energies are SFS quantities.

In analogy to the definition of the adiabatic index (with the subscript $s$ indicating constant entropy) 
\begin{equation}
    \gamma=\left( \frac{\partial \ln P}{\partial \ln \rho} \right)_s,
\end{equation}
we define $\geff$ as the full derivative of the pressure to density of a fixed parcel of gas along a Lagrangian path,
\begin{equation}
    \geff=\frac{d\ln P_{\rm tot}}{d\ln \rho}=\frac{\rho}{P_{\rm tot}}\frac{\dot P_{\rm tot}}{\dot \rho},
    \label{eq:geff}
\end{equation}
with the upper dot indicating full time derivative and $P_{\rm tot}$ as defined in eq.~(\ref{eq:ptot}).

\subsubsection{Ideal EoS with dissipation}
The time derivative of an ideal EoS (\equ{EoS}) can be separated into its isentropic and non-isentropic part. We first differentiate the pressure of a Lagrangian parcel of gas:
\begin{equation}
    \dot{P}=\left( \gamma-1 \right)\left( \dot{\epsilon}\rho+\epsilon\dot{\rho}\right),
    \label{eq:pdot}
\end{equation}
The time derivative of the specific energy, $\dot{\epsilon}$, can be taken from its thermodynamic definition,
\begin{equation}
    \dot{\epsilon}=-P\dot{V}-q,
    \label{eq:edot}
\end{equation}
with $V$ the specific volume ($V=\rho^{-1}$) and $q$ a general non-adiabatic energy sink rate. A negative value of $q$ corresponds to an energy source.
Inserting \equ{edot} into \equ{pdot} and using \equ{EoS} again, we find,
\begin{equation}
    \dot{P}=\gamma\frac{\dot{\rho}}{\rho}P-\left(\gamma-1\right)\rho q.
\end{equation}
The interaction between two forms of energy, such as the transfer between kinetic and magnetic components associated with dynamo action, can be incorporated into this framework by introducing a positive $q$ term into one form of energy (for example the kinetic), compensated by a negative contribution of equal magnitude to the other (for example magnetic energy).

\subsubsection{Kinetic EoS with dissipation}
Following \citet{maclow98} and \citet{robertson12} we model the dissipation rate of turbulence as proportional to the largest eddy turnover time (see \equ{dissip}), 
\begin{equation}
    q_{\rm dis}=\eta \frac{v}{a \lambda}\frac{v^2}{2}=\eta \frac{v^3}{a L},
    \label{eq:qdis}
\end{equation}
with $\eta$ a dimensionless free parameter (that depends on the numerical scheme and resolution) and $\lambda$ the largest eddy scale ($L/2$). When decay is efficient, we expect $\eta$ to be of order unity, but once a dissipationless flow pattern develops, it will be much smaller. This term dissipates kinetic into thermal energy, and formally should appear with a negative sign in the thermal component. However, by using an isothermal EoS for the gas, the thermal energy of the gas is fixed, and any heating of the gas is assumed to radiate out instantly. We simply introduce this term as cooling, directly from the kinetic pressure.

\subsubsection{Energy transfer between the kinetic and magnetic components}
Turbulence enhances initially small seeds of the magnetic field via small-scale dynamo processes \citep{BrandenburgSubramanian2005}. The amplification rate is initially exponential and eventually decreases to zero as the magnetic energy and the turbulent energy approach their saturation ratio, which is a function both of the Mach number and turbulent driving pattern \citep{federrath11} and the ratio of turbulent and magnetic dissipation, i.e, the magnetic Prandtl number \citep{SchekochihinEtAl2007,FederrathSchoberBovinoSchleicher2014,schober15}. Inspired by this behavior, we model the energy transfer rate from the kinetic to magnetic as
\begin{equation}
    \dot{\epsilon}_{\rm KB}=\Gamma \epsilon_{\rm B}\left(1-\frac{\epsilon_{\rm B}}{\sigsat\ekin}\right),
    \label{eq:ekb}
\end{equation}
where we recall that $\epsilon_{\rm B}$ and $\ekin$ are the sub-filter scale magnetic and kinetic energies. The saturation ratio, $\sigsat=\epsilon_{\rm B}^{\rm sat}/\ekin^{\rm sat},$ is the ratio of the two components before compression, and will be calibrated for each simulation depending on its setup. By analogy with the dissipation rate, the growth rate of the magnetic field, $\Gamma,$ is also taken as a fraction of the largest eddy turnover rate,
\begin{equation}
    \Gamma=\eta_{\rm B} \frac{v}{aL},
\end{equation}
with the dimensionless coefficient $\eta_{\rm B}$ of order unity.

While this internal energy transfer does not change the total energy content of the gas, it does change the total pressure, because of the difference in $\gamma$ for each component. We thus treat it as a sink term in $\dot{\epsilon}_{\rm kin}$ and as a source term in $\dot{\epsilon}_{\rm B}.$ We also note that this rate can become negative if the magnetic energy exceeds its saturation level with respect to the kinetic energy. While energy in such a physical case will flow from magnetic to kinetic, as the fields are strong enough to rearrange the material into a lower magnetic energy state, there is no justification to assume that the rate at which this happens is related to the eddy turnover rate. Regardless, we use \equ{ekb} even for that case. As we show later, this modeling, with the same coefficients for $\epsilon_{\rm B}\rightarrow \ekin$ as  $\ekin\rightarrow \epsilon_{\rm B},$ leads to reasonable results, although the value of $\eta_{\rm B}$ that we end up requiring is considerably less than unity. We leave further investigation into this point for later studies. 

\subsection{Calculation of $\geff$}
We have now everything in place for the calculation of $\geff.$
It is convenient to write the total pressure (eq.~\ref{eq:ptot}) as:
\begin{equation}
    P_{\rm tot}=P_{\rm th}\left(1+\alpha_k+\beta^{-1}\right),
\end{equation}
with 
\begin{align}
\alpha_k&=\frac{P_{\rm kin}}{P_{\rm th}}=\frac{1}{3}\frac{v^2}{c_s^2}=\frac{1}{3}\mathscr{M}^2,\nonumber\\
\beta&=\frac{P_{\rm th}}{P_{B}}={c_s^2\rho}\left(\frac{1}{3}\frac{B^2}{8\pi}\right)^{-1}\label{eq:alphabeta}
\end{align}
 with $\mathscr{M}$ the rms Mach number of the flow, and $\beta$ the plasma $\beta$ parameter for the coarse-grained case.

If the sound speed is constant as we have assumed, and we use the relation in \equ{pdot} for the kinetic and magnetic parts, we get:
\begin{equation}
\dot{P}_{\rm tot}=\frac{\dot{\rho}}{\rho}\left(P_{\rm th}+\frac{5}{3}P_{\rm kin}+\frac{4}{3}P_{\rm B}\right)-\frac{2}{3}\rho q_{\rm dis} - \frac{1}{3}\rho \dot{\epsilon}_{\rm KB}.
\end{equation}
By noting that (c.f., \equs{qdis}{ekb})
\begin{align}
\rho q_{\rm dis}&=3\eta\frac{v}{aL}P_{\rm kin},\nonumber\\
\rho\dot{\epsilon}_{\rm KB}&=3\eta_{\rm B}\frac{v}{aL} P_{\rm B}\left(1-\frac{\epsilon_{\rm B}}{\sigsat\ekin}\right),
\end{align}
and using the relation
\begin{equation}
\frac{\dot\rho}{\rho}=-3\frac{\dot a}{a},
\end{equation}
and \equ{alphabeta}, we get:
\begin{align}
\dot{P}_{\rm tot}&=\frac{\dot{\rho}}{\rho}P_{\rm th}\bigg[1+\alpha_k \left(\frac{5}{3}+\frac{2}{3}\eta\frac{v}{\dot{a}L} \right)\nonumber\\
&+\beta^{-1}\left(\frac{4}{3}+\frac{1}{3}\eta_{\rm B}\frac{v}{\dot{a}L}\left(1-\frac{\epsilon_{\rm B}}{\sigsat\ekin} \right) \right)\bigg] .\label{eq:pdot2}
\end{align}
Plugging \equ{pdot2} into \equ{geff}, we finally get:
\begin{align}
\geff&=\frac{1}{1+\alpha_k+\beta^{-1}}\bigg[ 1+\alpha_k \left(\frac{5}{3}+\frac{2}{3}\eta\frac{v}{\dot{a}L} \right)\nonumber\\
&+\beta^{-1}\left(\frac{4}{3}+\frac{1}{3}\eta_{\rm B}\frac{v}{\dot{a}L}\left(1-\frac{\epsilon_{\rm B}}{\sigsat\ekin} \right) \right)\bigg] .\label{eq:geff2}
\end{align}
Eq.~(\ref{eq:geff2}) manifests a few important physical properties. The equation properly interpolates between various extremes: when the thermal component is dominant over the kinetic and magnetic components, $1\gg\alpha_k,\beta^{-1}$, $\geff$ reduces to $1$, as expected (for an isothermal gas). Likewise, when the kinetic  component dominates, $\alpha_k\gg 1,\beta^{-1}$, $\geff$ reduces to $5/3$ (and to $4/3$ when the magnetic component dominates). The sign of the dissipation term and the energy transfer term depends on the sign of $\dot a.$  This is consistent with the expected behaviour that the dissipation always acts to reduce the pressure. When gas contracts ($\dot a <0$), $\geff$ drops, and the pressure growth is reduced. When gas expands ($\dot a >0$), $\geff$ increases, and the pressure drop due to the expansion, $P\propto \rho^{\geff}$ drops even faster because of the dissipation. The magnitude of the dissipation term is not theoretically bound, and $\geff$ can become negative or very large. This does not contradict our physical understanding. When large dissipation is present ($v\gg |\dot a|$), it is possible that as gas is compressed its pressure decreases, corresponding to a negative $\geff$.

\subsection{Comparison to simulations}
\label{sec:comparison}

In this section we analyze the reaction of multi-component gas to contraction, and compare the simulated runs (\se{sims} and \se{results}) to our analytic predictions. To quantify this reaction we use $\geff$ (\equ{geff}), and, in order to compare the simulations to our predictions, we derive this quantity in two separate ways. First, we take the logarithmic derivative of the total pressure $P_{\rm tot}$ (eqs.~\ref{eq:pkin_sim}, \ref{eq:pb_sim}, \ref{eq:ptot}) with respect to density directly from the simulation, $\gamma_{\rm sim}.$ Then, we compare it to our analytic prediction, $\gamma_{\rm pred},$ according to \equ{geff2}. 
It is of pedagogical and practical value to consider first a simplified version of our analytic model, for which zero dissipation is assumed. This is discussed in \se{gamma_nodissip}. A comparison to the full model is presented in \se{gamma_full}.

\subsubsection{$\geff$ comparison without dissipation}\label{sec:gamma_nodissip}

\begin{figure}
	\centering
	\includegraphics[width=0.475\textwidth]{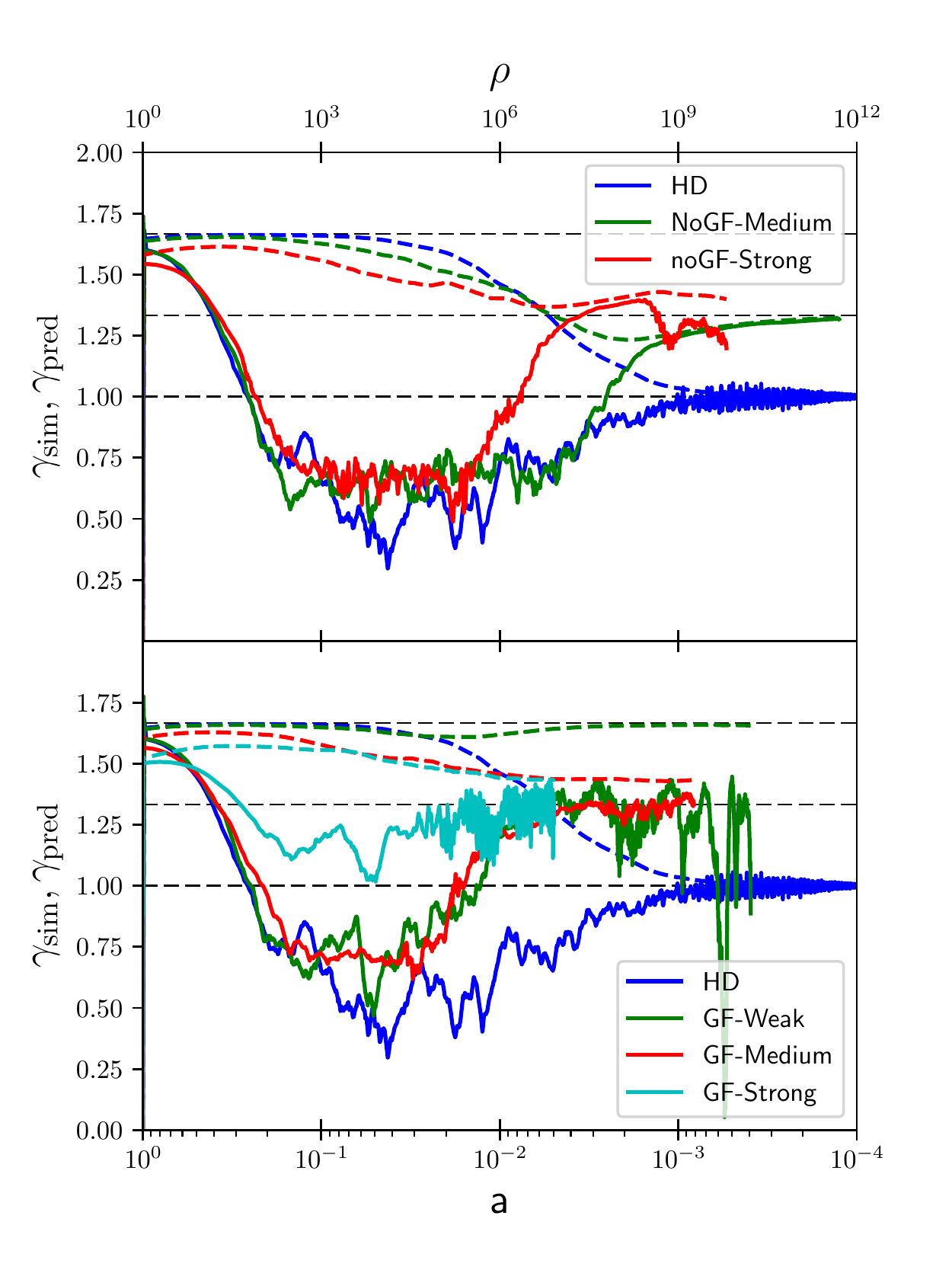}
	\caption{Calculated $\geff$ ($\gamma_{\rm sim},$ solid lines) and dissipationless predicted $\geff$ (\equ{geff2} with $\eta=\eta_{\rm B}=0$, dashed lines) as a function of scale factor and density, for the noGF runs (\top{}) and GF runs (\bot{}). The dashed horizontal lines mark the values of $1$, $4/3$, and $5/3$ that correspond to our expected values for thermal-, magnetic- and kinetic-dominated equations of state, respectively.}
	\label{fig:gamma_nodissip}
\end{figure}

\begin{figure}
	\centering
	\includegraphics[width=0.475\textwidth]{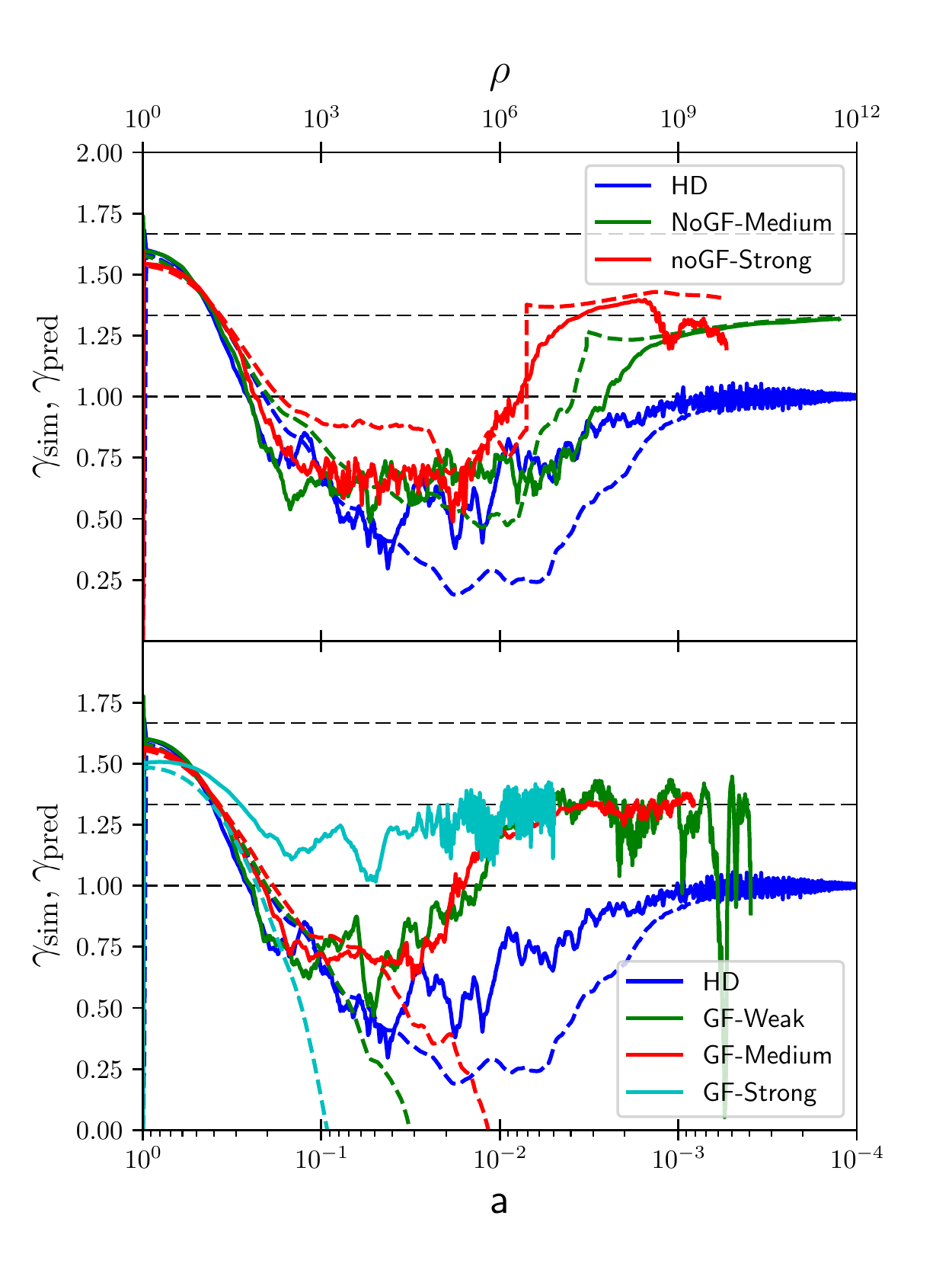}
	\caption{Same as \figref{gamma_nodissip}, but with the full predicted model (\equ{geff2} shown as dashed lines, which correspond to the expected behaviour in the presence of dissipation and kinetic-magnetic energy transfer where appropriate (see text).}
	\label{fig:gamma}
\end{figure}

Dissipation sinks energy from the gas as it is compressed. Therefore, we expect gas to be more compressible (i.e., lower $\geff$) when dissipation cannot be neglected. By contrast, if dissipation is weak enough, $\geff$ approaches the value expected for a gas with a mixture of thermal, turbulent, and magnetic pressure, with the different components weighted according to the relative pressures of each component. In terms of our model, this amounts to setting $\eta=\eta_{\rm B}=0$. As discussed in the previous sections, we expect this approximation to be valid at the initial stage, and then, to some degree, at late stages when a dissipationless flow (DLF) forms. 

\figref{gamma_nodissip} compares $\gamma_{\rm sim}$ (solid lines) with $\gamma_{\rm pred}$ (dashed lines). The horizontal dashed lines mark values of $\geff=1$, $4/3$, and $5/3$ that correspond to the expected behavior for purely isothermal, purely magnetic and purely kinetic gas, respectively. For the HD simulation, the gas initially starts close to the value predicted for supersonic turbulence, $\geff\sim5/3$, and as dissipation becomes dominant $\gamma_{\rm sim}$ drops to $\approx 1/2$. Once the turbulence has decayed significantly (\figref{mach_a} \top) and the kinetic pressure drops below the thermal pressure (\figref{beta_mach}), no energy is left to dissipate, and the gas behaves like an isothermal gas with $\geff=1$. The MHD runs without magnetic guide fields (noGF runs, \top{}) initially behave similarly, with $\geff$ dropping as dissipation becomes important. However, the dissipation stops for a different reason. At late times, after the transition to DLF, the gas behaves as magnetic field-dominated, with $\geff\approx 4/3$. The asymptotic behavior at the initial stage and then again for $a\lesssim 10^{-2}$ is well recovered by our analytic model without dissipation (\top{}, dashed lines), for the HD and noGF runs. Note that at $a\approx 1$ the predicted values are slightly higher than the simulated ones, indicating that some dissipation exists even for the very early stages when compression starts.

The \bot{} presents a similar calculation, but for the runs with guide fields (GF runs). In all three GF runs, $\gamma_{\rm sim}$ approaches $\geff\approx 4/3$ at large compressions. This occurs even when the magnetic pressure is sub-dominant to the kinetic one (GF-Weak) or comparable (GF-Medium and GF-Strong), and our model predicts that the gas behaves like a purely kinetic gas (GF-Weak, $\gamma_{\rm pred}\approx 5/3$) or intermediate (GF-Medium and GF-Strong, $4/3<\gamma_{\rm pred}<5/3$). This discrepancy indicates that some dissipation is present even at those late stages. In summary, the simplified non-dissipative model provides a very good prediction of the asymptotic values of $\gamma$ for the HD and noGF runs, and provides a reasonable prediction but with a slight overestimate for the asymptotic $\geff$ of the GF runs.  

\subsubsection{$\geff$ comparison of full model}\label{sec:gamma_full}

\figref{gamma} shows a similar plot to \figref{gamma_nodissip}, but with $\gamma_{\rm pred}$ that includes dissipation as the dashed lines. First examine the top panel. In this plot, we have used $\eta=1.8$ and $\eta_{\rm B}=0.01$ at compressions $a \gtrsim 10^{-2}$. These are by-eye fits; a more systematic calibration is certainly possible, but since it is likely that $\eta$ depends on the numerical scheme and resolution, and $\eta_{\rm B}$ depends on the specific geometry of the driving, we find little practical value in estimating them more robustly here. To obtain a meaningful physical results, we would need to explicitly include physical dissipation terms, i.e., kinematic viscosity and magnetic resistivity \citep{federrath11,FederrathSchoberBovinoSchleicher2014,Federrath2016jpp}. 

For our purpose, it suffices to show that we can fit the numerical results with reasonable values of $\eta$ and $\eta_{\rm B}$. The inclusion of the kinetic dissipation term, $\eta v/aL$ improves the small mismatch at $\rho\approx 1$ that was present in \figref{gamma_nodissip}. This illustrates that for our simulation setup, dissipation makes a minor difference from the beginning of the compression phase, even though it was not directly observed in \figref{beta_mach}. For the HD and noGF runs (\top{}), the predicted $\geff$ curve is a reasonably good fit to the simulated one, even when the dissipation dominates.

However, the fit would fail miserably when flows enter the self-avoiding channel-flow phase. At that stage, our model would predict an ever-increasing dissipation because of the rising turbulent velocity when, in fact, very little dissipation takes place. Similarly, as the magnetic energy exceeds its saturation level, more and more energy would be predicted by our model to be pumped from magnetic to kinetic when, in fact, little energy does. To mimic this drop in transfer terms, we simply set $\eta = \eta_{\rm B} = 0$ when $(\alpha_k\beta)^{-1}=P_{\rm B}/P_{\rm kin}>3$. In the Figure, the shutdown of the transfer terms produces the discontinuity in the predicted $\geff.$ Without it, $\geff$ would first shoot up because of the large $\dot{\epsilon}_{\rm KB}$, and then shoot down because of the large $q_{\rm dis}.$ Obviously, the transition to the ineffective dissipation regime is not sharp, and a better fit between the calculated and predicted models can be obtained with a more sophisticated transition model. However, at this stage, we value the simplicity of the model over its accuracy. The success of the dissipation model when $P_{\rm B}\lesssim P_{\rm kin},$ combined with its failure beyond that point and the success of the dissipationless model there, is the key physical finding of this paper.

Our model is somewhat less successful in the case when a guide field is present, as shown in the lower panel of \figref{gamma}. The model is reasonable at first, but, as noted above, the switch to DLF appears to depend not just on the ratio of magnetic to kinetic pressures, but also on the field topology. Our switch at $P_{\rm B}/P_{\rm kin}>3$ is too conservative for this case, and as a result predicts strong dissipation at intermediate values of $a$ when in fact our simulations are already switching to a low-dissipation state. We could improve the fits by hand-tuning when we switch to $\eta = \eta_{\rm B} = 0$, but without an understanding of exactly how this switch depends on the magnetic topology, we would have to do so independently for each run, which seems of little value.

Regardless of our ability to predict exactly when the switch to DLF will occur, we can still give a physical interpretation to our results as follows. With our simulation setup, the turbulence at the compression stage is driven by globally re-normalizing the velocity and distance. This type of driving enhances existing modes, and allows the system to settle into ``natural modes'' which are not forced by an arbitrary external driving.  From our results it appears that even a relatively small decay of some of the modes (e.g., even in the case of a sub-dominant magnetic field) is enough to significantly decrease the power in these modes and allows the flow to settle into a DLF.

\section{Implications for ISM}\label{sec:discussion}

While the study of dissipation in scale-free compression of turbulence described above is general, our motivation for conducting this study is not. We seek to find how magnetic fields alter the collapse of GMCs and, in particular, to test whether realistic initial conditions may delay the dissipation and consequent collapse of the GMCs. The ISM, in which GMCs form, is a multi-phase medium, with subsonic turbulence in the diffuse, warm and hot phases and supersonic turbulence in the denser, cold phases such as GMCs. The entire ISM is furthermore interlaced with magnetic fields and immersed in cosmic rays and radiation fields. The turbulence is maintained by various energy sources, including driving by feedback from star formation, gravitational collapse, and galaxy dynamics \citep{FederrathEtAl2016,FederrathEtAl2017iaus}. In this work we simplify this system to manageable levels by focusing on its most basic aspects: we start with gas that is supersonic and magnetized and test its response to quick compression. The compression pumps energy into the turbulence and into the magnetic field on all scales, without imposing any particular scale or randomness through external driving. We find that even weak magnetic fields eventually force the gas to settle into a channel-flow pattern with greatly-reduced dissipation. This is in stark contrast to the case where magnetic fields are absent, where we find that (in agreement with previous purely hydrodynamical simulations -- \citealt{robertson12}) that the kinetic energy steadily decays and dissipation remains significant throughout the entire evolution of the gas.

The greatly reduced dissipation rate for the flow that we find in the presence of a magnetic field indicates that significantly less energy input may be required to produce the observed linewidths in GMCs than had previously been conjectured. However, a full exploration of this issue will need to take into account many more processes, among them the multiphase nature of the ISM and realistic stellar feedback. The latter process, which will drive random turbulence on a typical scale corresponding to the distance between stars within each GMC, may actually increase the dissipation rate by disturbing the dissipationless flow. Thus, while stellar feedback is highly energetic, it could prove to be an effective cooling agent by increasing the dissipation. Even within our toy model, a more systematic test of various compression models and rates is needed to test the universality (or lack-thereof) of the dissipation model and the calibrated rates. We leave these tests to future work.

At a minimum, however, we note that our results suggest that dissipation inside a forming GMC will be much less than has commonly been assumed. The gas from which GMCs form is observed to be threaded by a significant net magnetic flux \citep{LiHenning2011,LiEtAl2011,PillaiEtAl2015}, and in this case the flow can be nearly dissipationless almost immediately after compression begins. Even in the limiting case of zero net flux but fields close to equipartition, as observed, the dissipation rate is substantially reduced once the cloud has compressed by a factor of $\sim 100$ in linear dimension. This is not all that much by interstellar standards: the mean density of the Milky Way's ISM is $n\sim 1$ cm$^{-3}$, so a factor of 100 linear compression corresponds to a density $n\sim 10^6$ cm$^{-3}$, i.e., typical of a prestellar core.

The onset of dissipationless flow will not necessarily halt collapse. Even in the most favorable case, when no dissipation occurs, highly magnetized gas is unstable to scale-free collapse because its effective EoS is $\geff=4/3$, which is right at the critical value for hydrostatic gravitational stability ($\gamma_{\rm crit}=4/3$). This indicates that, as a cloud compresses, the force exerted by the magnetic field outwards grows just as fast as the gravitational forces increases inwards. This general point has been overlooked in the past, because the focus was on comparing timescales rather than using the formal requirement for hydrostatic atmospheres. However, we note that if the external compression is filamentary as is often found, the critical value for stability drops to $\geff=1,$ and, if linear magnetic fields prevent compression perpendicular to the magnetic-field direction, the collapse can only occur along one direction, for which the critical value is $\geff=0$ \citep{birnboim16}.

The compression rate of observed GMCs can be characterized in a way that relates it to our ideal simulations. We define the non-dimensional compression rate  $H_{\mathrm{ND}}=-\tsound H$ which is our (absolute value of the) Hubble coefficient for the compression in units of sound crossing time. If compression is driven by gravitational collapse, this non-dimensional compression rate is (up to order of unity corrections) $\tsound/\tff,$ with $\tff$ the gravitational free-fall time. By replacing $\tsound$ of our turbulent cloud by $\mach\tturb$ (which is, again, correct up to order of unity corrections) we get $H_{\mathrm{ND}}=\mach\tturb/\tff$. The ratio of the turbulent timescale to the free fall timescale for reasonable GMC's is $\tturb/\tff=2/\sqrt{\alpha_{\rm vir}},$ with $\alpha_{\rm vir}\approx 1,$ the virial parameter for virialized GMCs \citep[see sec. 8.3 of][]{krumholz17}, yielding $H_{\mathrm{ND}}\approx 2\mach.$
In our numerical simulations $\tsound=1$ and $H_{\mathrm{ND}}=-H$ spanning a range between $1$ and $200$ (see Table~\ref{tab:sims} and Appendix~\se{H1}). Since typical Mach numbers for local GMCs \citep{schneider13} and in the Central Molecular Zone Cloud G0.253+0.016 \citep{federrath16a} are $\mach\approx 10$ and for high-z GMCs or ULIRGS can be as high as $\mach\approx 100,$ we argue that our simulations bracket the observed range. Furthermore, as gas is compressed, dissipation always dominates eventually over the adiabatic compression. Faster compression rate simply extends the initial stage in which dissipation can be neglected. We note, however, that unlike realistic GMCs, our Mach number is independent of the compression rate and is set by the initial driving. Our Mach numbers ($\mach \approx 10$) are smaller than is observed for ULIRGS, and is perhaps comparable to that of local GMCs. Additionally, it does not follow the correlation set by $H_{\mathrm{ND}}\approx 2\mach$ that is expected by observations. Since the onset of the dissipation is predicted by comparing the dissipation timescale (ie. turbulent timescale) and compression timescale, we expect that realistic GMCs will always start near the onset of the dissipation phase, without the long adiabatic phase seen for $H=-200$ in our simulations. We leave a more systematic test of the dependence of our conclusion on the compression rate and Mach numbers for future work.

\section{Summary and conclusions} \label{sec:conclusions}

In this paper, we study the evolution of magnetized supersonically turbulent gas as it is compressed. The compression is scale-free and corresponds to gravitational compression that operates on all scales, much like the expansion of the universe in cosmology, but with a scale factor that decreases with time (negative Hubble constant). Our simulations of this scale-free compression are performed by using a modified version of the cosmological expansion model in the FLASH MHD code (\S\ref{sec:sims}), and explore a range of magnetic field strengths and net fluxes.

The scale-free compression enhances all turbulent and magnetic modes by the same factor, and does not impose any arbitrary scale or randomness of phases. Consequently, the system is allowed to relax into a self-consistent state for which naturally decaying modes decay away while non-decaying modes are enhanced by the compression. We find that this relaxation, combined with a magnetic field, produces a surprising result: after some time the gas re-arranges itself into a self-avoiding channel flow, in which state the dissipation rate is nearly zero. This occurs whether or not there is a net magnetic flux, but the transition happens more readily for non-zero magnetic flux and for stronger fields, and it does not occur at all in the absence of a magnetic field.

We interpret the simulations by comparing them to a theoretical model for coarse-grained MHD turbulence. Our model treats the flow has having three distinct energy reservoirs (thermal, kinetic, and magnetic) that are coupled by dynamo action and dissipation. This model allows us to construct an equation of state with an effective adiabatic index $\geff$, whose value depends on the relative balance between the different energy reservoirs and on the overall rate of dissipation. We calibrate the transfer and dissipation terms from the simulations, and show that, once calibrated, the model provides a good match to our numerical experiments. A key features of the model, and of the numerical results it describes, is that once the flow is sufficiently magnetically-dominated the dissipation rate for the flow is nearly zero, and compression drives a continual increase in the kinetic energy per unit mass and the Mach number.

The existence of this dissipationless state may have significant implications for the ISM, and for giant molecular clouds in particular. These clouds are assembled by compression of regions of atomic ISM with significant magnetic flux. If this compression resembles the idealized gravitational contraction we consider here, then GMCs may be in flow states where the dissipation rate is much less than is commonly assumed based on earlier work in non-compressing regions and on non-magnetized flows. This in turn might significantly ease the problem of how GMCs maintain their large linewidths while still forming stars with very low efficiencies.

Finally, we note that the analytic model derived here constitutes a first step towards a physically-motivated sub-resolution model for the ISM. Given some idea \citep[perhaps calibrated by observations; see][]{birnboim15} about the content of the turbulence, thermodynamics and magnetic fields of the ISM, our model predicts the behavior of the gas in a way that can be implemented into large-scale, low-resolution simulations that only resolve the ISM as a coarse-grained mixture of the components. This too will be investigated in future work.

\section*{acknowledgements}

We thank Chalence Safranek-Shrader and Romain Teyssier for their help during the implementation of the Hubble source terms for MHD (\S\ref{sec:sims}). Y.B.~wishes to thank the Research School of Astronomy and Astrophysics at The Australian National University for hosting him on a sabbatical during 2016--2017. 
C.F.~gratefully acknowledges funding provided by the Australian Research Council's Discovery Projects (grants~DP150104329 and~DP170100603). M.~R.~K.'s work was supported under the Australian Research Council's \textit{Discovery Projects} funding scheme (project DP160100695). The simulations presented in this work used high performance computing resources provided by the Leibniz Rechenzentrum and the Gauss Centre for Supercomputing (grants~pr32lo, pr48pi and GCS Large-scale project~10391), the Partnership for Advanced Computing in Europe (PRACE grant pr89mu), the Australian National Computational Infrastructure (grant~ek9), and the Pawsey Supercomputing Centre with funding from the Australian Government and the Government of Western Australia, in the framework of the National Computational Merit Allocation Scheme and the ANU Allocation Scheme. 
The simulation software FLASH was in part developed by the DOE-supported Flash Center for Computational Science at the University of Chicago.

\bibliographystyle{mnras}
%%\bibliography{yuval,federrath,krumholz}

\appendix
\section{Numerical tests}\label{sec:numerics}
\subsection{The $\nabla\cdot B=0$ constraint}\label{sec:divB}
\begin{figure}
	\centering
	\includegraphics[width=0.475\textwidth]{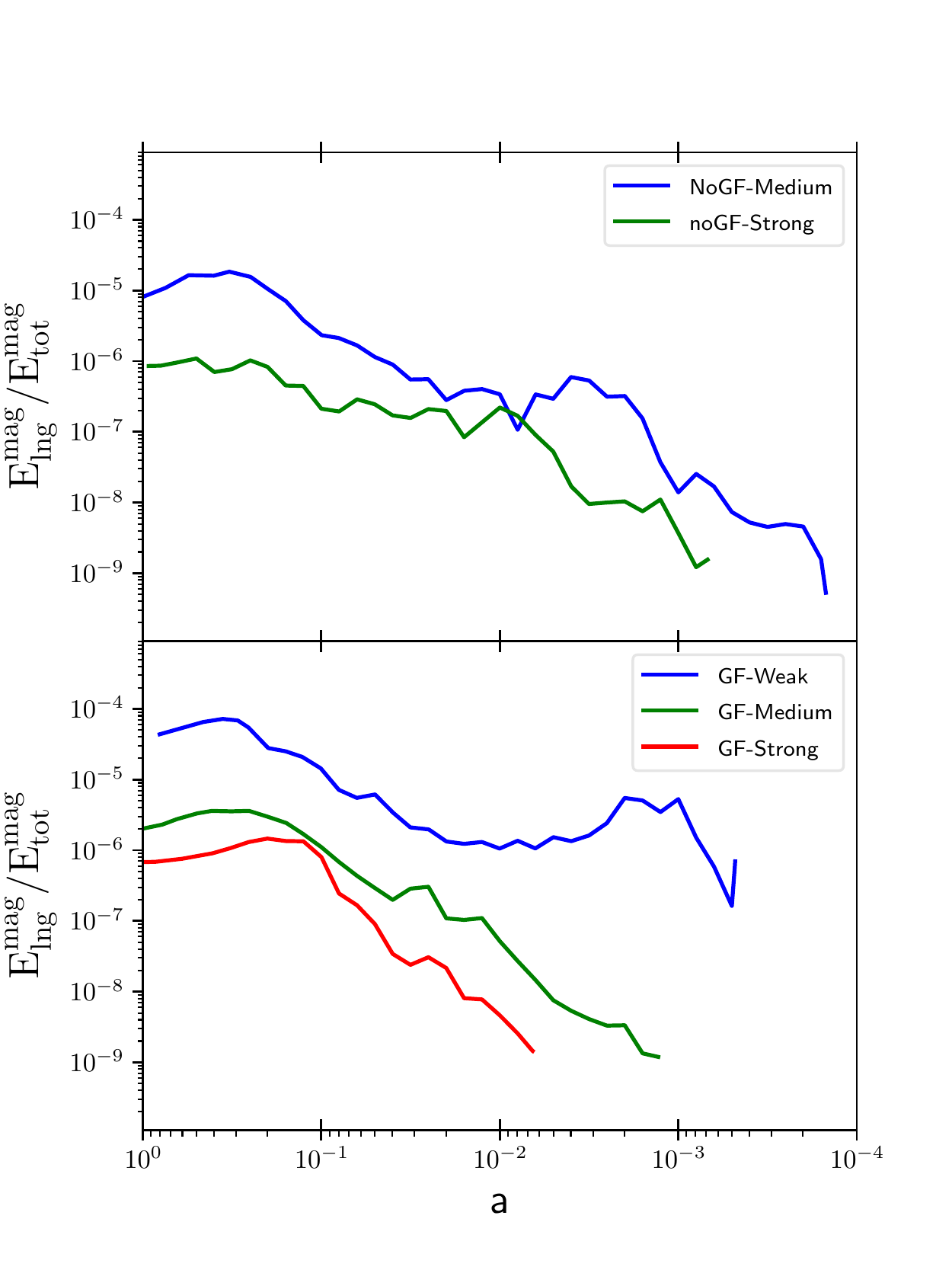}
	\caption{The ratio of longitudinal (div B) modes to the total magnetic energy (solenoidal + longitudinal modes) for our simulation suite. The relative energy in div(B) modes stays below $10^{-4}$ for all times and all scale factors, and is thus negligible.}
	\label{fig:divB}
\end{figure}
Physically, no energy should propagate into longitudinal modes of the magnetic field. For highly supersonic, low plasma $\beta$ turbulence simulations in FLASH using the HLL3R and HLL5R solver \citep{WaaganFederrathKlingenberg2011} this has been demonstrated and compared with alternative schemes in \citet{waagan11}. However, the compression in the simulations presented here could, in principle, change this conclusion because the amplitude of the magnetic fields is enhanced due to adiabatic compression.
\figref{divB} presents the ratio of magnetic field energy in longitudinal modes to the total energy in magnetic fields. As is evident, this value is never larger than $10^{-4}$, and, except for GF-Weak, smaller than $10^{-5}$. Additionally, this value drops as compression occurs, and is largest near the comencement of compression. We therefore do not expect the numerical errors in $\nabla\cdot B$ to effect the conclusions of this paper.

\subsection{Convergence with numerical grid resolution}\label{sec:convergence}
\begin{figure}
	\centering
	\includegraphics[width=0.475\textwidth]{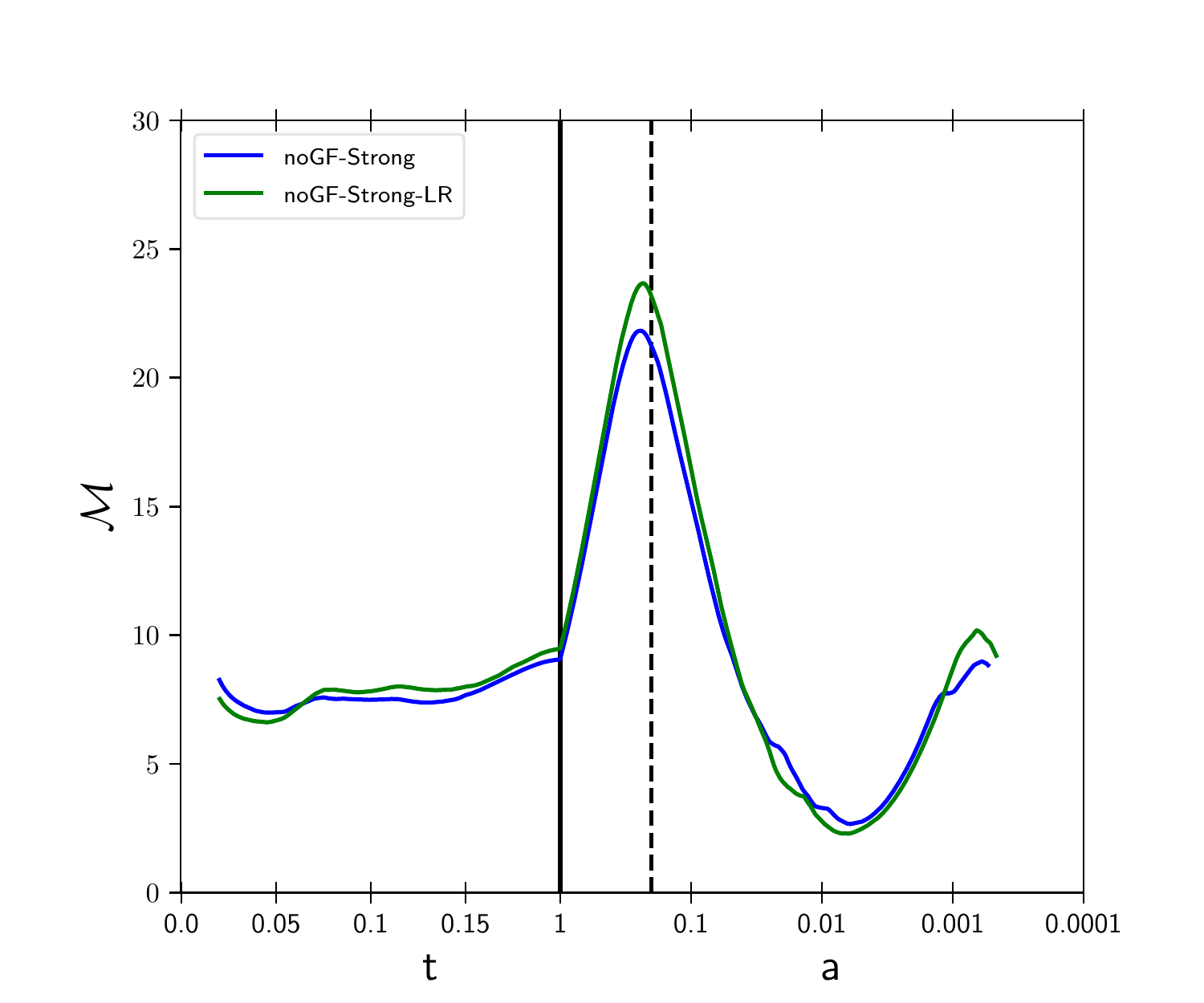}
	\caption{The Mach number dependence of time and compression (see \figref{mach_a} and its legend) for our noGF-Strong and noGF-Strong-LR, which share the same physical setup, but with the LR simulation with half the resolution along each axis.}
	\label{fig:M_at_converge}
\end{figure}
\begin{figure}
	\centering
	\includegraphics[width=0.475\textwidth]{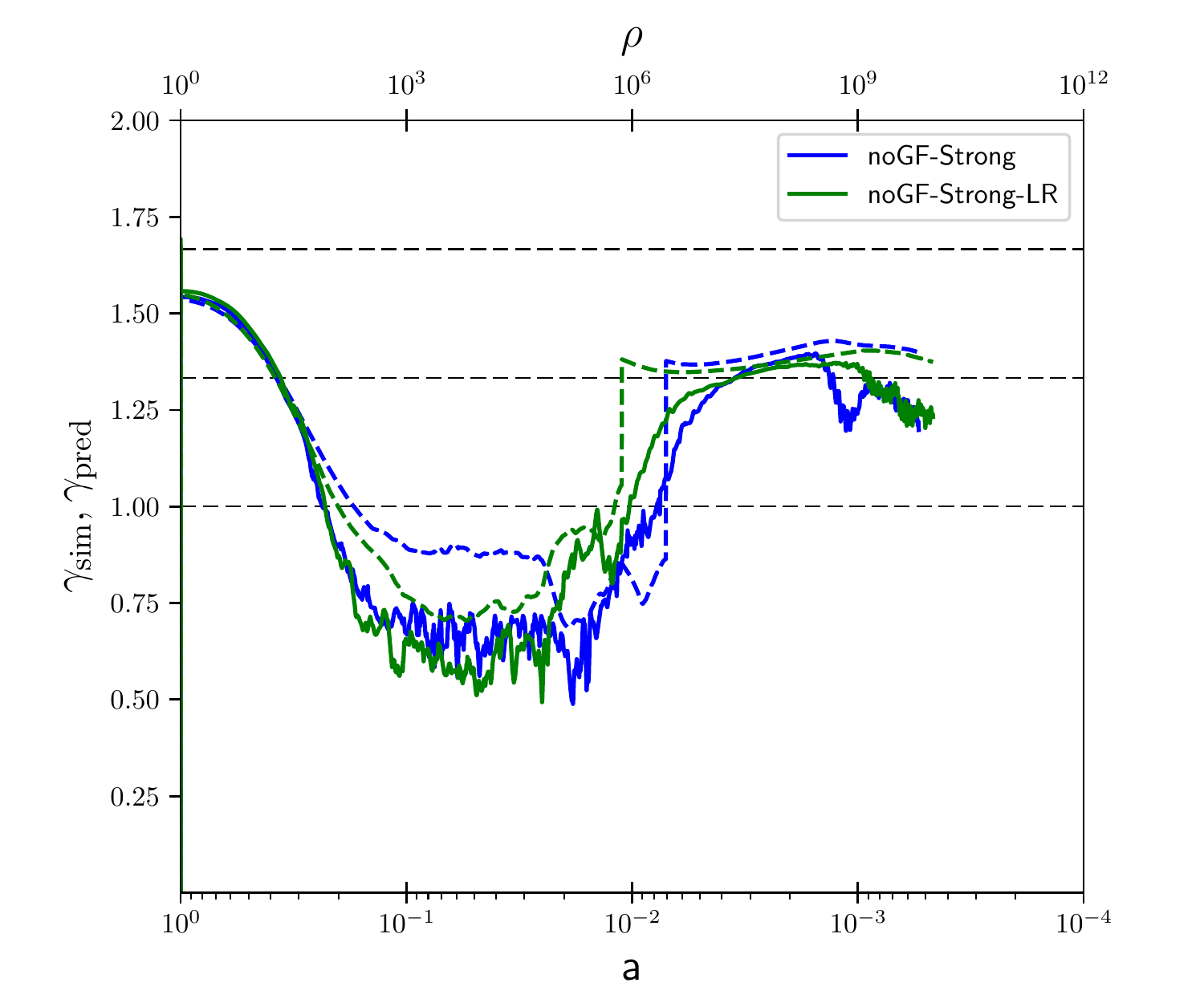}
	\caption{The simulated and predicted $\geff$ of the nominal resolution run (noGF-Strong) and low resolution one (noGF-Strong-LR) with full dissipative terms (see \figref{gamma} and its legend).}
	\label{fig:gam_converge}
\end{figure}
It is well known that the necessary grid resolution for simulating fully-developed turbulence with an inertial subrange is at least $1024^{3}$ cells \citep[e.g.,][]{KleinEtAl2007,KritsukEtAl2007,SchmidtEtAl2009,FederrathDuvalKlessenSchmidtMacLow2010,Federrath2013}. However, this was not our goal in this paper. We expect that for our particular focus on coarse-grained values averaged over large scales, such a high resolution is not critical. In the interest of computational efficiency and to allow for many different simulations, we only ran $512^{3}$-cell boxes. Indeed, previous studies have demonstrated that large-scale averages converge even at a resolution of $256^3$ grid cells \citep{KitsionasEtAl2009,FederrathDuvalKlessenSchmidtMacLow2010,PriceFederrath2010,KritsukEtAl2011Codes}. In this appendix we briefly demonstrate that our resolution of $512^{3}$ cells is sufficient for our needs.

We check for convergence by comparing two physically similar MHD runs: noGF-Strong and noGF-Strong-LR (see \cref{tab:sims}). The two simulations differ only in that the the former has a lower resolution of $256^{3}$, while the latter has our standard $512^3$ resolution. We find that, at the end of the initial driving stage, the Mach numbers for the two runs differ by $5\%$, while the saturation levels for the magnetic field strength differ by about $30\%$. While these differences are not negligible, they do not change the qualitative results once compression begins. We demonstrate this in \figref{M_at_converge}, which shows the Mach number evolution, and \figref{gam_converge}, which shows $\geff$; these figures can be compared to \figref{mach_a} and \figref{gamma} in the main text. We see that the overall behaviour of both the Mach number and the adiabatic index are nearly identical in the two runs.

\section{Slow ($H=-1$) compression}\label{sec:H1}
\begin{figure}
	\centering
	\includegraphics[width=0.475\textwidth]{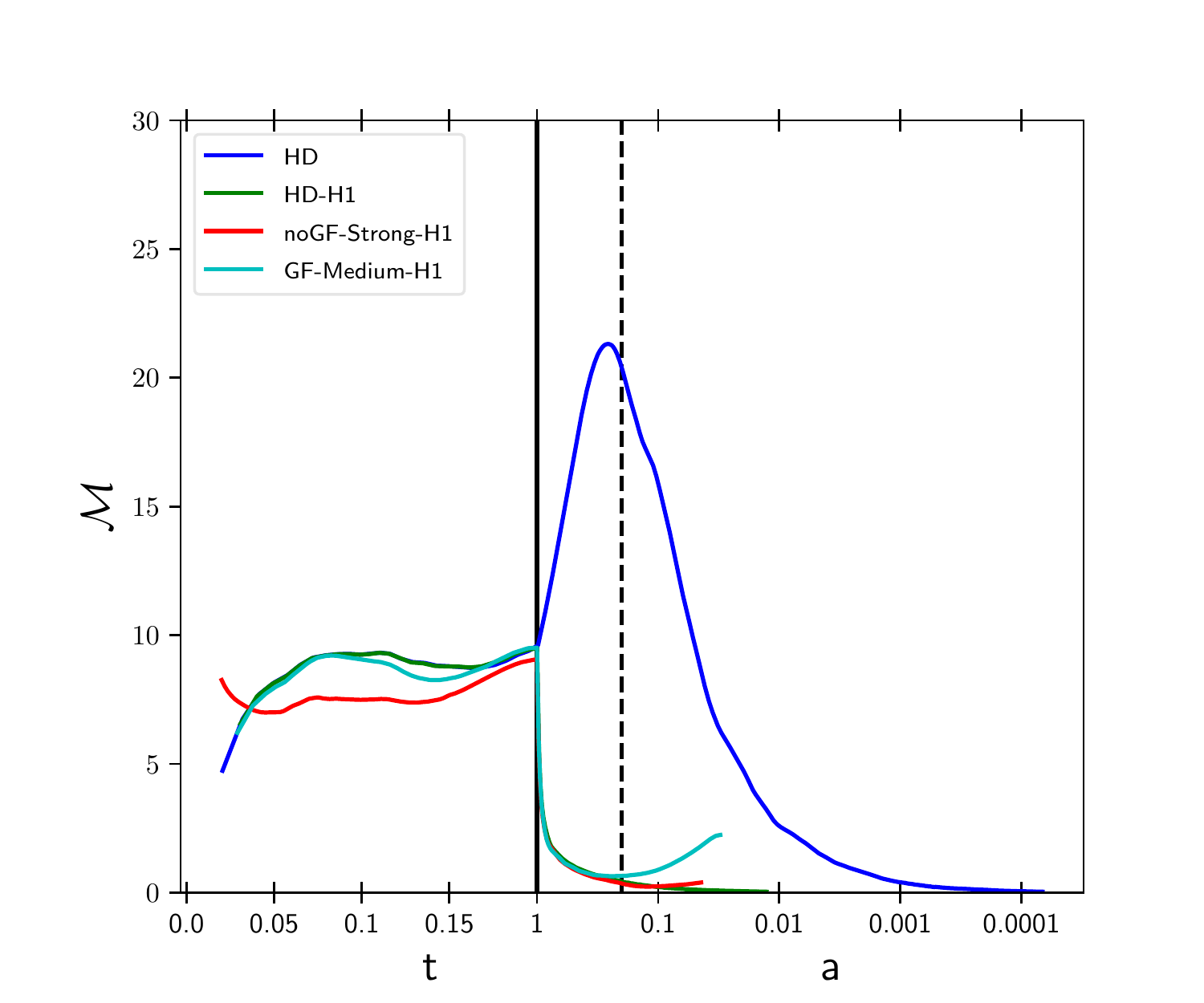}
	\caption{Similar to \figref{mach_a}, but presenting the evolution of the average Mach number for slow compression runs.}
	\label{fig:M_at_H1}
\end{figure}
\begin{figure}
	\centering
	\includegraphics[width=0.475\textwidth]{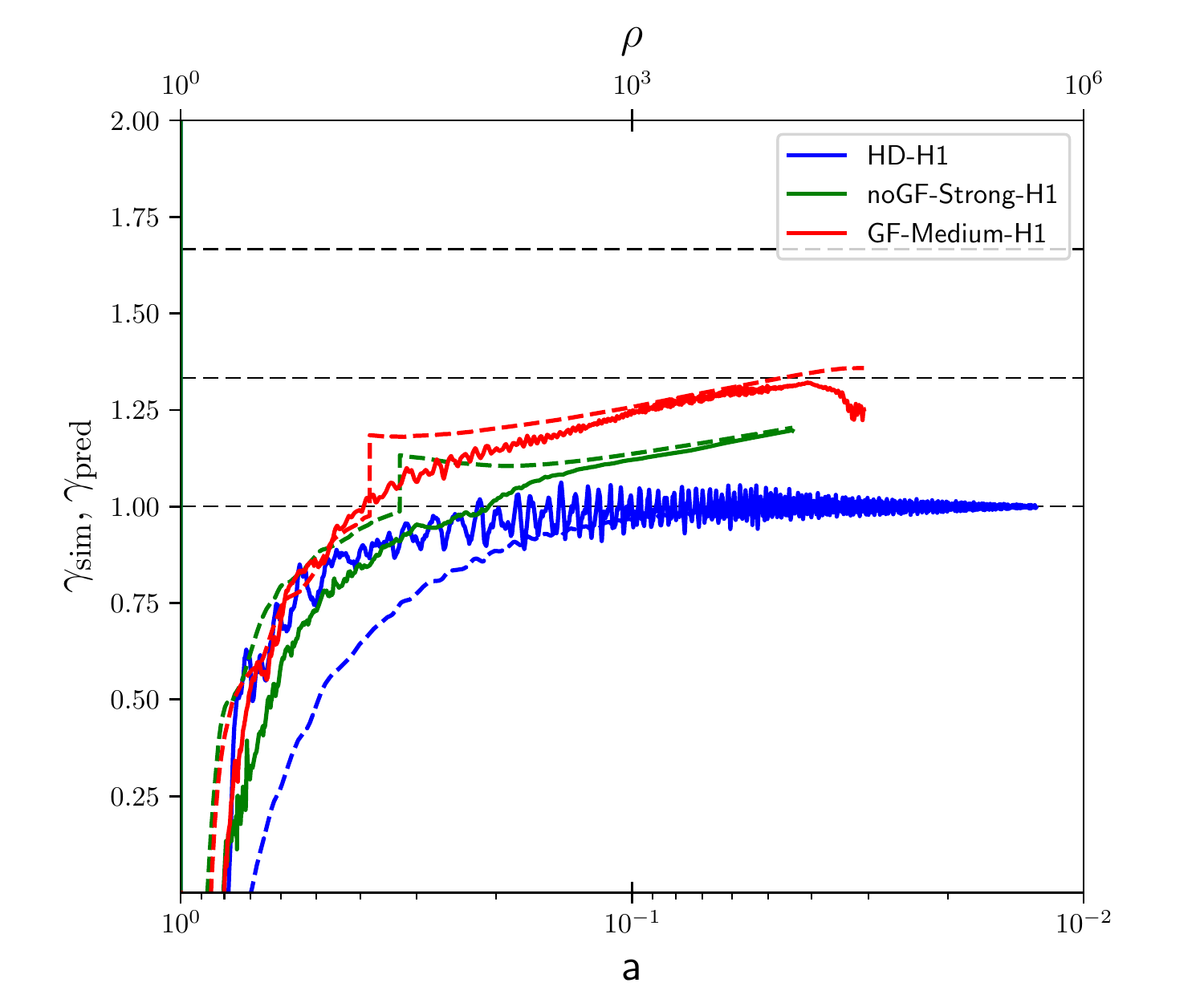}
	\caption{Similar to \figref{gamma}, but presenting the simulated and predicted $\geff$ for the slow compression runs.}
	\label{fig:gam_H1}
\end{figure}

All the simulations used in the main text have $H=-200$ corresponding to very fast compression. The motivation for this choice is that, given our high Mach numbers ($\mathcal{M} \approx 10$, motivated by the properties of observed molecular clouds), the dissipation timescale is very short. Consequently, values of $H$ near unity would not show a distinct amplification stage before the onset of compression, and instead would proceed directly to the dissipation stage. We find that ensuring the presence of a distinct amplification phase helps elucidate the physics of the problem, which is why we elected to use $H=-200$ as our standard choice. However, it is important to demonstrate that our central result, the onset of dissipationless flow, is independent of this choice. For this reason, in \figref{M_at_H1} we show the evolution of the Mach number for three simulations (HD-H1, noGF-Strong-H1 and GF-Medium-H1; see \cref{tab:sims}) with a compression rate $H=-1,$ as well as our fiducial HD run to guide the eye. This compression rate corresponds to the gas compression at roughly the sound speed, significantly slower than what would be expected if the gas were to contract at free-fall.

In these runs, the Mach number declines immediately once driving is turned off and compression starts, as we would expect for such slow compression. However, in these runs we still see the characteristic increase in Mach number at late times that occurs due to the onset of dissipationless flow. We demonstrate this more clearly in \figref{gam_H1}, which shows $\geff$ measured for the $H=-1$ simulations, as compared to our theoretical model using the same values of $\eta$ and $\eta_{\rm B}$ (including the condition when we set these terms to zero) calibrated from the $H=-200$ simulations, and used for \figref{gamma} in the main text. We first note that, as in the $H=-200$ simulations, at late times the MHD runs have $\geff$ substantially above unity, demonstrating the reduced dissipation that is the central result of this work. Moreover, the plot shows that our model continues to provide a very good description of the value of $\geff$. Since dissipation is dominant from the beginning, our model predicts the initial value of $\geff$ to be very low, as is the case. At late times, our model predicts that $\geff$ should increase as energy is pumped into the magnetic component, and that eventually dissipation should turn off, leading to an increase in $\geff$. Both of these predictions are borne out by the simulations. The success of our model's $\gamma_{\rm pred}$ at reproducing the value $\gamma_{\rm sim}$ we measure from the simulations, using the same calibration of $\eta$ and $\eta_{\rm B}$, suggests that our results are not dependent on the choice of a particular compression rate $H$. 

\label{lastpage}

\end{document}